# Numerical study of friction reduction and underlying mechanism of air film via parallel and tandem injections on a hypersonic flat plate at high altitude

Xiuzheng Cheng[a*], Mengyu Wang[a*], Qin Li[a†], Lingyun Wang[a], Yihui Weng[a], Pan Yan[a]

a. Xiamen University, China, 361102

**Abstract:** Friction force reduction is crucial to enhancing the lift-to-drag ratio of hypersonic vehicles. As a flow control technique, air film offers efficient and robust drag reduction under extreme conditions. This study uses numerical simulations to analyze a flat-plate with air films generated via parallel and tandem injections under freestream conditions of $Ma_\infty = 15$ at an altitude of $H = 60 km$. The numerical approaches are validated before performing the grid convergence study. Simulations are then carried out using various mass flow rates and jet spacing, followed by an examination of the effect of the streamwise-to-spanwise aspect ratio of the injection hole. The cross-section characteristics and flow topology are then analyzed, the drag reduction and associated costs are evaluated, and the underlying mechanisms are explored, yielding the following key findings: (1) Under hypersonic and high-altitude conditions, the air injection lifts the incoming flow, producing a pronounced low-speed region near the injection site and over the downstream wall. A plateau appears in the wall-normal distribution of the streamwise velocity, with a value significantly lower than that of its freestream counterpart. This reduces the near-wall velocity gradient, thereby achieving drag reduction. However, the subsequent flow reattachment may lead to localized drag increases; (2) As the mass flow rate of injection increases, the wall-normal extent of the near-wall low-speed layer expands and the film coverage broadens; however, the increased friction limits gains in drag reduction, and greater flow losses are also incurred; (3) Increasing the hole spacing in the parallel injection reduces the penetration height of the film, the spanwise coverage, and the drag reduction efficiency, as well as decreasing the flow losses. In contrast, for the studied tandem injections, no significant differences are observed with analogous spacing variations; (4) Increasing the injection aspect ratio enlarges the spanwise coverage of the low-speed layer, improves the spanwise uniformity of the streamwise velocity distribution, diminishes the friction-increasing zones, and significantly enhances the drag reduction. Notably, the consequences are the inverse of those of the low-speed situation.

**Keywords:** hypersonic friction reduction; air film; mass flow rate; injection spacing; aspect ratio of hole

## 1 Introduction

Minimizing aerodynamic drag is a fundamental approach to improve the lift-to-drag ratio of hypersonic vehicles at high altitude. Drag consists of wave drag and friction drag. Advancements in design technology have continuously reduced wave drag, increasing the prominence of friction drag. It has been reported that friction drag can account for up to 90% of the total drag under extreme conditions [1]. Therefore, it is crucial to develop reliable and efficient drag reduction techniques for high-speed vehicle design [2].

Surface friction can be effectively reduced using flow control techniques [3], which are generally divided into two categories: passive control and active control [4]. Although extensive research has been conducted on these techniques, including notable achievements and applications, several limitations and challenges remain. For instance, passive controls, such as aerodynamic spikes [5] and forward-facing cavities [6][8], may not fully meet the demands of hypersonic vehicles as they suffer from issues including ablation, the induction of unsteady flows, unstable drag reduction, and the potential for structural damage under hypersonic conditions. Active controls, such as the opposite jet [9]-[12][12], also tend to induce unsteady flows. However, air film, a specific active control, shows promise under extreme conditions. This technique operates by injecting working gas through slots or holes at a certain angle to form a thin gaseous layer across the surface that effectively separates the external flow from the wall. Studies by Chang et al. [13] have demonstrated that such an injection can not only lower boundary layer temperatures but also reduce viscosity and shear stress within the boundary layer, thereby achieving both drag reduction and thermal protection. Although air film (or film cooling) technology has been widely used in aero-engines [14], Zuo et al.'s [15] review noted that research on integrating air film for both thermal protection and drag reduction remains insufficient, with most existing studies focusing on internal flows such as aircraft channels, leaving external flows comparatively understudied.

---

[*] Xiuzheng Cheng and Mengyu Wang are co-first authors
[†] Corresponding author, email: qin-li@vip.tom.com

Extensive studies have been conducted on air film in recent years, examining factors such as blowing ratio, inter-hole spacing, and hole geometry. Depending on the inclination of the angle of the jet to the freestream direction, the configurations can be broadly categorized as either tangential or inclined/vertical injections, as illustrated in Fig. 1. The air film is typically generated using two methods: discrete holes and slot/groove. Although there are extensive studies on low-speed situations, this work specifically focuses on hypersonic scenarios. With respect to tangential film injection, Chen et al. [16] numerically studied the cooling effectiveness of the slot method and compared films over flat and curved surfaces at $Ma_\infty = 6.0$, showing that those over curved surfaces exhibited superior cooling performance. Hu et al. [17] numerically investigated the flow and corresponding heat transfer in a nozzle under an air-based tangential film generated by slot at $Ma_\infty = 6.0$, demonstrating that the film significantly reduced the wall heat flux at the throat with minimal influence on the velocity profile at the exit. Mo et al. [18] numerically studied a tangential air film generated by slot over a hypersonic cone at $Ma_\infty = 15.0$, showing that increasing the slot height and injection mass flow rate improved lift-to-drag ratio; however, the benefit of increasing slot height was limited regarding drag reduction. Shifting the slot location forward was found to enhance drag reduction but potentially reduce the lift. Lin et al. [19][20] conducted experiments in a $Ma_\infty = 3.8$ wind tunnel to study tangential film cooling over a supersonic cone. They observed a decrease in the scope of effective cooling on the windward side with increasing angle of attack, while the cooling effectiveness on the leeward side increased correspondingly. Zhao et al. [21] investigated a tangential slot injection over a supersonic cone at $Ma_\infty = 6.0$, finding that the film could significantly reduce the drag despite a relatively high friction near the injection exit. Additionally, Sun et al. [22][23] studied a $Ma_\infty = 2.35$ slot-injected air film on hypersonic optical domes in the $Ma_\infty = 7.1$ wind tunnel and found that reducing the slot height and increasing the jet Mach number and the jet-to-freestream static pressure ratio could effectively enhance film cooling.

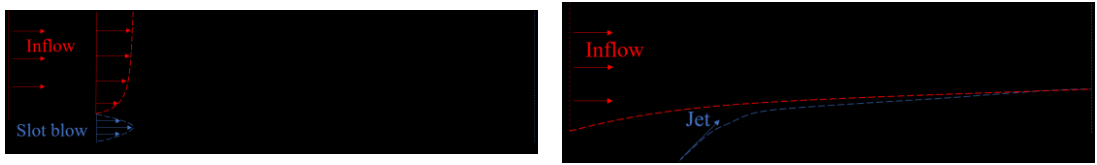

(a) Tangential jet                                      (b) Inclined jet

Fig. 1: Schematic diagram of air film generated by two types of injections.

Regarding air films generated by inclined jet, Hombsch et al. [24] investigated the effects of the inclination angle of the slot on a flat plate under different Reynolds numbers ($Re$), Mach numbers, and flow types (laminar and turbulent). Pudsey et al. [25] used simulations under $Ma_\infty = 4.5$, finding that drag reduction performance was enhanced by decreasing the spacing between injection holes and increasing the mass flow rate. Combining simulations and experiments, Keller et al. [26] investigated the film cooling of various gases under $Ma_\infty = 2.6$. Their results revealed that gases with low molar mass, such as helium and hydrogen, exhibited a superior drag-reduction effect. Zhou [27] examined the flow characteristics and performance of film cooling with discrete holes on a flat plate with $Ma_\infty < 2.4$, including cylindrical, merged, and twin holes. Kerth et al. [28] investigated the mechanisms of the porous injection of nitrogen and helium in a 7.5° cone at $Ma_\infty = 7.1$, finding that the helium injection led to a greater reduction of heat flux. Furthermore, Dai et al. [29] studied film injection through discrete holes on an aircraft rudder at $Ma_\infty = 6.0$ and discovered that for vertical injections, hole placement upstream weakened thermal protection. Conversely, reducing the jet-wall inclination angle enhanced coolant adherence. Xiang et al. [30] noted in their review that the study of porous film cooling and hole geometry is still needed. It is noteworthy that Mao et al. [31] recently numerically studied film cooling by discrete micro-holes at $Ma_\infty = 13.2$ on a flat plate, considering over 100,000 staggered micro-holes (each 0.254 mm in diameter), differing from the usual discrete-hole configurations with a finite scale.

Despite the above progress in research on air film control under supersonic and hypersonic conditions, several limitations remain to be addressed: (1) Although high-speed air film techniques have been successfully applied in areas such as aero-engines, the applicability of parameters including blowing ratio, injection arrangement, and hole geometry to hypersonic scenarios remain insufficiently understood; existing studies have primarily emphasized thermal protection, paying relatively less attention to friction reduction; (2) Most current investigations focus on freestreams with $Ma_\infty$ less than 10, whereas those concerning high altitude conditions and higher $Ma_\infty$ are relatively scarce. Some studies implement tangential films via slot injection, which may create a potential disadvantage in terms of the structural strength of a vehicle; (3) Research on film cooling by discrete injection have primarily been confined to two-dimensional or single-hole configurations, with less exploration of multi-holes and their corresponding three-dimensional (3-D) flow interactions; (4) The mechanisms underlying friction reduction by air film under hypersonic conditions remain inadequately understood, e.g., those related to discrete holes with a finite scale. Most existing work focuses on evaluating the overall friction reduction efficiency but lacks detailed analyses of the flow physics and reduction mechanisms; for instance, the mechanisms revealed by the present study, whereby jet injection may lead to increased local friction and how a velocity plateau forms in the wall-normal distribution of the streamwise velocity. Given that real gas effects become pronounced at high altitudes (e.g., 50–70 km) and high $Ma_\infty$, high-fidelity numerical investigations that incorporate corresponding effects are undoubtedly important when investigating friction reduction under such extreme conditions.

Given the above issues, this study employs an equilibrium gas model to conduct systematic simulations and analyses of air films generated by hole injections on a flat plate in a hypersonic flow. In particular, it focuses on high-altitude hypersonic conditions and discrete injections with inclined angles, finite size, and in parallel and tandem layouts from perspectives such as the flow structures, friction reduction performance, and underlying mechanisms. Furthermore, it examines the influence of injection hole shape on friction reduction. The structure of this paper is as follows: Section 2 introduces the governing equations, the equilibrium gas model, the WENO3-$Z_{ES3}$ scheme [32] for high resolution simulations, and their validations; Section 3 presents a grid convergence study, followed by simulations of parallel injection under various mass flow rates and hole spacings, subsequently comparing flow structures and friction reduction; Section 4 conducts similar investigations of tandem injections; Section 5 focuses on the effects of the aspect ratio of injection holes on the friction reduction; Section 6, presents the conclusions that can be drawn.

## 2 Numerical methods

### 2.1 Governing equations

To model the hypersonic flow over a flat plate at high altitude under conditions of $Ma_\infty = 15$ and $H = 60 km$, this study uses the 3-D laminar Navier–Stokes equations; their forms in Cartesian coordinates can be expressed as follows:

$$\frac{\partial Q}{\partial t} + \frac{\partial (E-E_v)}{\partial x} + \frac{\partial (F-F_v)}{\partial y} + \frac{\partial (G-G_v)}{\partial z} = 0, \quad (1)$$

where $Q$ is the vector of conservative variables; $E, F$, and $G$ are the inviscid fluxes in the $x, y$, and $z$ coordinates, respectively; and $E_v, F_v$, and $G_v$ are the corresponding viscous fluxes. Their explicit forms are illustrated as follows.

$$Q = \begin{bmatrix} \rho \\ \rho u \\ \rho v \\ \rho w \\ \rho e \end{bmatrix}, E = \begin{bmatrix} \rho u \\ \rho u^2 + p \\ \rho uv \\ \rho uw \\ (\rho e + p)u \end{bmatrix}, F = \begin{bmatrix} \rho v \\ \rho uv \\ \rho v^2 + p \\ \rho vw \\ (\rho e + p)v \end{bmatrix}, G = \begin{bmatrix} \rho w \\ \rho uw \\ \rho vw \\ \rho w^2 + p \\ (\rho e + p)w \end{bmatrix}$$

$$E_v = \begin{bmatrix} 0 \\ \tau_{xx} \\ \tau_{xy} \\ \tau_{xz} \\ u\tau_{xx} + v\tau_{xy} + w\tau_{xz} - q_x \end{bmatrix}, F_v = \begin{bmatrix} 0 \\ \tau_{yx} \\ \tau_{yy} \\ \tau_{yz} \\ u\tau_{yx} + v\tau_{yy} + w\tau_{yz} - q_y \end{bmatrix}, G_v = \begin{bmatrix} 0 \\ \tau_{zx} \\ \tau_{zy} \\ \tau_{zz} \\ u\tau_{zx} + v\tau_{zy} + w\tau_{zz} - q_z \end{bmatrix}.$$

where $q$ is the heat flux and $\tau$ is the viscous stress. Further explanations of the conventional terms are omitted in the above formulations, while those regarding the equilibrium gas model will be explained in Section 2.2.

### 2.2 Equilibrium gas model

The ideal gas assumption becomes invalid under hypersonic flow conditions due to significant real gas effects. To account for these influences, this study adopts the equilibrium gas model (EGM). Specifically, the curve-fitting method, proposed by Srinivasan and improved by Tannehill et al. (1987) [33], is employed, offering advantages in terms of both programming simplicity and computational efficiency. The EGM represents the thermodynamic properties and transport coefficients as piecewise functions, each defined over specific intervals. Each segment is constructed using Grabau-type transition functions with the following form:

$$f(x,y) = f_1(x,y) + \frac{f_2(x,y) - f_1(x,y)}{1 \pm exp(k_0 + k_1 x + k_2 y + k_3 xy)}. \quad (2)$$

In the above expression, the sign "+" in the denominator denotes an odd function transition, while "−" corresponds to an even function transition. The coefficients $k_i$ are constants. The functions $f_1(x,y)$ and $f_2(x,y)$ are cubic polynomials, expressed as follows:

$$f_1(x,y) = p_1 + p_2 x + p_3 y + p_4 xy + p_5 x^2 + p_6 y^2 + p_7 x^2 y + p_8 xy^2 + p_9 x^3 + p_{10} y^3, \quad (3)$$
$$f_2(x,y) - f_1(x,y) = p_{11} + p_{12} x + p_{13} y + p_{14} xy + p_{15} x^2$$
$$+ p_{16} y^2 + p_{17} x^2 y + p_{18} xy^2 + p_{19} x^3 + p_{20} y^3,$$

where $p_i$ are constant coefficients.

Using Eq. (2), the temperature $T$, as the thermodynamic representative, is fitted and expressed as $T = T(e, \rho)$ or $T = T(p, \rho)$, and the equivalent specific heat ratio $\tilde{\gamma}$ is formulated as $\tilde{\gamma} = \tilde{\gamma}(p, \rho)$. Based on these relationships, the enthalpy is given by $h = (p/\rho)(\tilde{\gamma}/\tilde{\gamma} - 1)$, and the speed of sound is expressed as $a = \sqrt{\tilde{\gamma} R T}$.

In the above fitting, the quantities are transformed into logarithmic form. Taking $T = T(p, \rho)$ as an example, the fitting expression is given by:

$$\log_{10} \frac{T}{T_0} = f(Y, Z), \quad (4)$$

where $X = \log_{10}(p/p_0)$, $Y = \log_{10}(\rho/\rho_0)$, $Z = X - Y$, $T_0 = 273.15K$, $\rho_0 = 1.292 kg/m^3$, $p_0 = 1.0134 \times 10^5 N/m^2$.

To determine transport properties, dynamic viscosity $\mu$ can be fitted as a function of either $\rho$ and $T$, or $\rho$ and $e$. The Prandtl number $Pr$ is fitted using $\rho$ and $T$, while thermal conductivity $k$ is fitted using $\rho$ and $e$. When $e$ and $\rho$ are used as variables, a logarithmic transformation is applied during the fitting of transport coefficients. For example, the fitting of dynamic viscosity as a function of density and temperature, $\mu = \mu(\rho, T)$, is given by:

$$\mu/\mu_0 = f(X, Y), \quad (5)$$

where $X = T/1000K$, $Y = \log_{10}(\rho/\rho_0)$, $\mu_0 = 16.5273 \times 1.058 \times 10^{-6} kg \cdot s/m$, and $\rho_0 = 1.243 kg/m^3$. Details of the formulations can be found in [33].

### 2.3 WENO3-$Z_{ES3}$ to discretize derivatives of inviscid flux

To facilitate the introduction of the numerical method, a brief preliminary description is provided. As illustrated in [32], consider the one-dimensional hyperbolic conservation law $u_t + f(u)_x = 0$ regarding the function $f(u)$ of $u$, supposing $\partial f(u)/\partial u > 0$. The conservative approximation of $f(u)_x$ at $x_j$ can be expressed as $\frac{\partial f(u)}{\partial x_j} \approx \frac{(\hat{f}_{j+1/2} - \hat{f}_{j-1/2})}{\Delta x}$, where $\hat{f}_{j+1/2}$ denotes some numerical flux. Assuming that $r$ is the number of sub-stencils and the grid number of each stencil, WENO-JS schemes with an order of $r$ can be formulated as $\hat{f}_{j+1/2} = \sum_{k=0}^{r-1} \omega_k q_k^r$ with $q_k^r = \sum_{l=0}^{r-1} a_{kl}^r f(u_{j-r+k+l+1})$, where $q_k^r$ is the candidate scheme with coefficients $a_{kl}^r$, and $\omega_k$ is the normalized nonlinear weight corresponding to the linear counterpart $d_k^r$. $\omega_k$ is typically derived from the non-normalized weight $\alpha_k$; for the WENO-Z type scheme, $\alpha_k$ may take the form $\alpha_k = d_k \left(1 + c_\alpha \left(\frac{\tau}{\beta_k^{(r)} + \varepsilon}\right)^p\right)$, where $\beta_k^{(r)}$ and $\tau$ represent the local and global smoothness indicators respectively, $p = 1$ or 2, $c_\alpha$ may take 1, and $\varepsilon$ can take a small value such as $10^{-40}$. The related denotations are suggested in [32], where a detailed accuracy analysis is conducted on the third-order WENO scheme (where $r = 2$) and a novel critical point accuracy analysis is proposed for computing the derivatives of the inviscid fluxes of Eq. (1). Unlike existing studies which typically assume the critical point located at $x_j$, this approach considers the critical point spanning the scheme's dependent stencils. Based on this understanding, it was demonstrated [32] that the classical three-point stencil $\{x_{j-1}, x_j, x_{j+1}\}$ fails to achieve the optimal order recovery at the first-order critical point.

Furthermore, by incorporating sufficient conditions to acquire the optimal order, a new global smoothness indicator was derived as $\tau_4$. To enhance the numerical robustness, the local smoothness indicator $\beta_k$ was revised. As a result, the nonlinear weights $\alpha_k$ in WENO3-Z$_{ES3}$ are formulated by choosing $c_\alpha = 0.4$, $\varepsilon = 1 \times 10^{-40}$, the revision of $\tau$ as $\tau_4 = |(2f_{j+1} - 3f_j + f_{j-1})(f_{j+2} - 3f_{j+1} + 3f_j - f_{j-1})|$, and the modified $\beta_k^{(2)}$ as $\{\beta_0^{(2)*} = \beta_0^{(2)} + c_{\beta_0}(f_j - 2f_{j-1} + f_{j-2})^2, \beta_1^{(2)*} = \beta_1^{(2)} + c_{\beta_1}(f_{j+2} - 2f_{j+1} + f_j)^2\}$ with $C_{\beta 0} = 0.5, C_{\beta 1} = 0.15$. Further details of the scheme can be found in [32].

**2.4 Validation tests**

Two cases were tested to verify the above EGM and WENO3-Z$_{ES3}$ scheme: the case of hypersonic flow over a cone and the case of a hollow cylinder with extended flare (HCEF). In addition to the employment of the above scheme and model, the computations employed the Steger–Warming flux vector splitting scheme.

(1) Hypersonic flow over a cone at $Ma_\infty$ =25.3

The other inflow conditions include $Re/m = 1.29 \times 10^5/m$ and $T_\infty = 252.6K$; and a grid is used with the number 200×100. The distributions of $u/U_\infty$ at $x = 1m$ along the $y$ coordinate and the distribution of the wall friction coefficient $C_f$ are presented in Fig. 2, which compares them with those from the 3-D upwind PNS (3-D_UPS) code developed by Tannehill et al [33]. The distributions currently predicted by the EGM corroborate those by [33]. In addition, the figure provides the results from the ideal gas model, which exhibit distinct deviations from those of the EGM, indicating that it is necessary to consider the real gas effect.

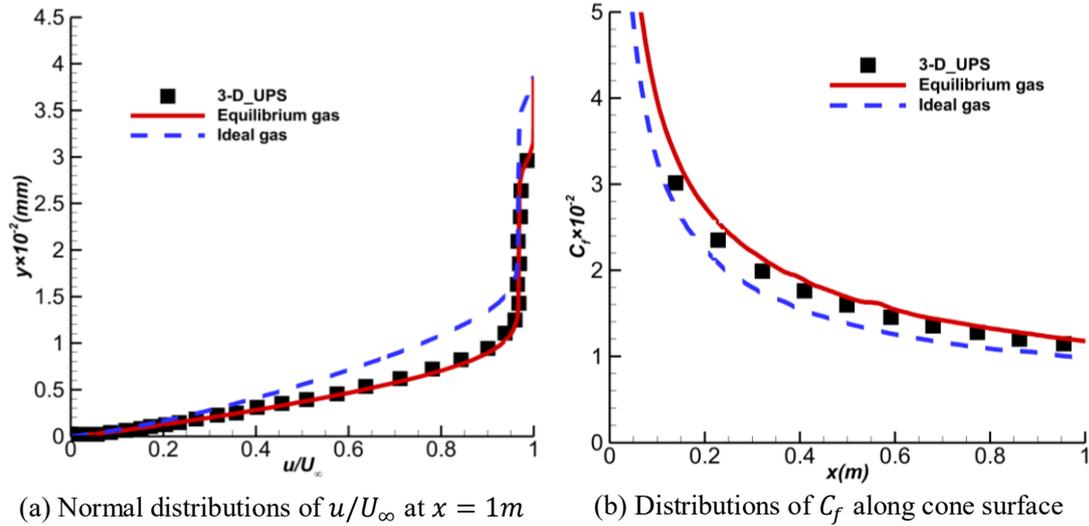

(a) Normal distributions of $u/U_\infty$ at $x = 1m$   (b) Distributions of $C_f$ along cone surface

Fig. 2: Variable distributions of hypersonic cone flow at $Ma_\infty$ =25.3.

(2) Hypersonic flow around a hollow cylinder with extended flare at $Ma_\infty$ =11.35

The other inflow conditions were set as follows: $Re/m = 3.596 \times 10^5/m$, $T_\infty = 79K$, and $T_w = 294K$. To demonstrate the performance disparity, WENO3-JS is also employed to produce surface pressure coefficients $C_p$ for comparison with those produced by WENO3-Z$_{ES3}$, with the results presented in Fig. 3. The figure shows that WENO3-Z$_{ES3}$ provides significantly better agreement with the experimental data than WENO3-JS.

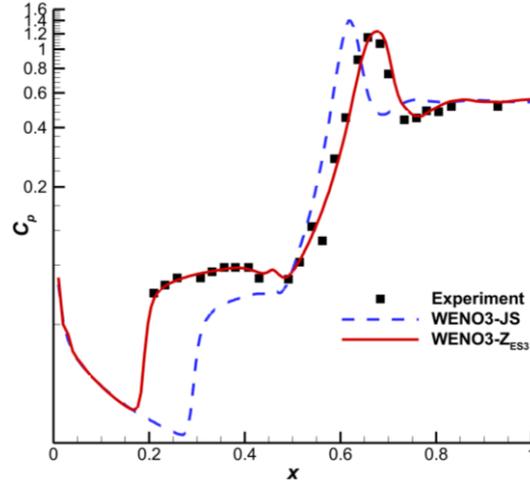

Fig. 3: Distributions of $C_p$ along HCEF surface at $Ma_\infty = 11.35$.

The above results demonstrate the successful validation and verification of the physical models and numerical methods employed in this study.

## 3 Flow structures and friction reduction in parallel injections under various conditions

As noted in the introduction, research on the characteristics of air films generated by parallel injection under hypersonic conditions remains limited, and understanding of the influence factors and underlying mechanisms is therefore insufficient. The present study indicates that friction reduction is influenced not only by injection $\dot{m}$ (mass flow rate) and $L_z$ (spacing between parallel injection holes), but is also closely related to the aspect ratio of the injection hole. To investigate the effect of aspect ratio, a series of configurations of injection hole—JET1, 2, and 3—are designed with different aspect ratios (details provided in Section 5). In this section, JET2 is selected to represent simulations under different mass flow rates and inter-hole spacings. The corresponding effects are analyzed, and the subsequent friction reduction efficiencies are computed and assessed, along with further analysis to elucidate the mechanism underlying the friction reduction.

### 3.1 Setup of parallel injections and grid convergence studies

#### 3.1.1 Freestream conditions, geometry of injection holes, and study plan

This study considered a flat plate at an altitude of 60 km under the following inflow conditions: $Ma_\infty = 15.0$, $Re_\infty/m = 9.04 \times 10^4/m$, $T_\infty = 255.78K$, and $T_w = 500K$. Fig. 4 presents the top and front views of the configuration of flat plate with parallel injection holes for this study. The total length of the plate is $L = 800mm$, and both the spanwise width and the spacing between injection holes equal $2L_z$. Considering the spanwise periodicity, the actual computational domain is represented by a solid line in the plane view, with the injection hole centered in the spanwise direction at $L_x = 250mm$ downstream from the leading edge of plate. The hole has a depth of $L_d = 6.45mm$ and an elliptical outlet on the horizontal plane has a spanwise major axis with a length of $D$. The Mach number at the hole inlet is set as $Ma_{jet} = 0.6$ with a total temperature $T_0 = 300.0K$. As shown in Fig. 4, the adjacent virtual injection holes are symmetrical with the solid lines $z = \pm L_z$ in the $xz$ plane. On this basis, the left boundary is defined as a supersonic inflow, while the right and upper boundaries are set as outflows; symmetry boundaries are chosen on both spanwise sides.

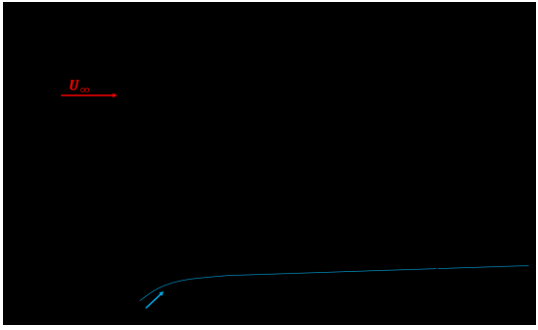 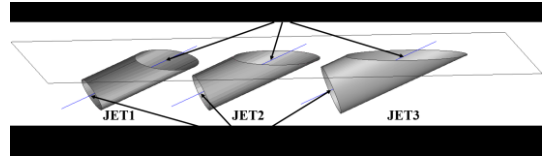

| Fig. 4: Flat plate with parallel injection holes (Top: Plan view, Bottom: Front view). | Fig. 5: Schematics of injection holes with different streamwise-to-spanwise aspect ratios. |

For the convenience of subsequent discussions, Fig. 5 presents a complete description of three injection holes (JET1-3). The centerlines of all holes form an angle of $\theta = 25°$ with respect to the flat plate. Specifically, the $D$ of JET2 is 3.5 $mm$, and the streamwise-to-spanwise ratio of the hole outlet's major axis is defined as AR; therefore, JET2 has an AR of 2:1. As shown in the figure, all holes (with straight walls) have inlets in the shape of ellipses formed by a perpendicular intersection with the hole's centerline. Keeping the area of the outlet (and the inlet) constant and shortening AR to 1:1 yields JET1, while elongating it to 4:1 produces JET3. JET2 is selected as the study object in this section. Simulations are conducted to investigate the effects of $\dot{m}$ and $L_z$ using two plans. In the first plan, $L_z = 15mm$ and $\dot{m} = 0.5, 1.0, 1.5, 2.0 g/s$ (with corresponding blow ratios of 38.3, 76.6, 114.9, 153.2) and in the second plan, $\dot{m} = 1.5 g/s$ and $L_z = 15, 25, 35, 45 mm$. Next, the grid convergence study is conducted.

### 3.1.2 Grid convergence study

The present study selected the conditions $L_z = 45mm$ and $\dot{m} = 0.5 g/s$ for JET2. Three structured grids are generated: coarse (1 million cells), medium (2 million cells), and fine (4 million cells) grids. The normal spacing of the first grid layer is $5 \times 10^{-5} m$ for the three grids. Fig. 6 illustrates the grid structures near the injection hole for the medium grids. The grid convergence study is conducted under the conditions specified in Section 3.1.1.

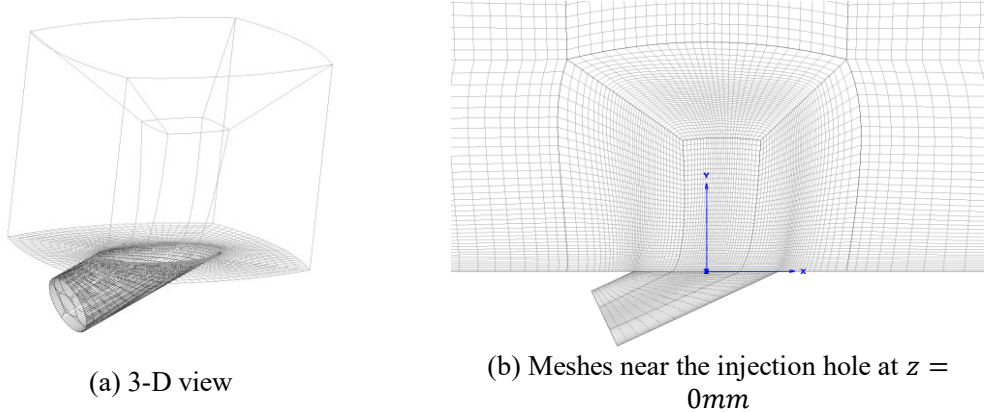

| (a) 3-D view | (b) Meshes near the injection hole at $z = 0mm$ |

Fig. 6: Schematic diagram of meshes around the injection hole.

Table 1 presents the drag coefficients $C_D$ for the flat plate across the three grids. The results show that the coarse grid yields noticeable deviations when compared with the medium and fine grids, the results of which are in close agreement.

Table 1: Wall drag coefficients of different grids.

| Mesh | Coarse | Medium | Fine |
|---|---|---|---|
| $C_D \times 10^3$ | 5.87 | 5.93 | 5.94 |

Fig. 7 presents the distributions of $C_p$ along the centerline of three grids, demonstrating that all distributions agree well with each other upstream of the injection, with the distribution of the coarse grid differing from those of the fine and medium grids downstream, indicating that the medium grid has achieved grid convergence. To balance accuracy and efficiency, the medium grid is selected for subsequent computations.

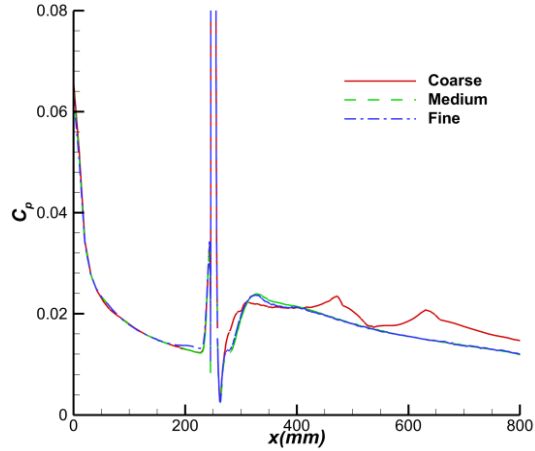

Fig. 7: Distributions of $C_p$ along $x$ direction across three grids.

### 3.2 Flow structures under different conditions

Representative cases are studied to explore the flow structures with different $\dot{m}$ and $L_z$. This section presents and discusses the streamwise velocity $u$ and streamline distributions on the streamwise cross-section for these cases, as well as the isobar contours and streamlines on the wall and the spanwise symmetry plane ($z = 0$).

#### 3.2.1 Patterns under different mass flow rates

Two representative conditions, $\dot{m} = 0.5,\ 2.0 g/s$, are selected to investigate the flow structures under different injection mass flow rates. Fig. 8 presents the contours of $u$ and streamline distributions of various streamwise cross-sections, with purple lines indicating 3-D streamlines. The results reveal that: (1) The cross-sectional streamlines indicate the presence of a pair of counter-rotating vortices. The vortices rotate inwardly near the injection holes, while downstream their rotation direction reverses. Because of the application of symmetric boundary conditions in the spanwise direction, the streamlines run parallel to the spanwise boundary; (2) Increasing the mass flow rate ($\dot{m}$) enlarges the downstream region (visualized in blue) of low $u$, and the incoming flow (in red) is lifted to a higher position by the air film. This is conducive to enhancing friction reduction; (3) Some 3-D streamlines originating from the injection holes pass through the downstream vortex cores, suggesting that symmetric vortex formation partially stems from the injection flow; (4) For the case where $\dot{m} = 0.5 g/s$, the vortices are closer to the wall and smaller in scale. The location where $(y, z) = (0, 0)$ at the cross-section downstream of the injection primarily exhibits a saddle point, corresponding to separation streamlines on the wall, indicating the existence of a relatively simple separation structure in the cross-section. For the $\dot{m} = 2.0 g/s$ condition, the vortices are farther from the wall and larger in scale. In the region where $z > 0$, the primary vortex rotates counterclockwise when observed along the positive $x$ axis, which is the opposite of the rotation direction when $\dot{m} = 0.5 g/s$. The location where $(y, z) = (0, 0)$ in the downstream cross-section forms a node; the saddle point connected to it suggests the onset of more complex 3-D flow separation.

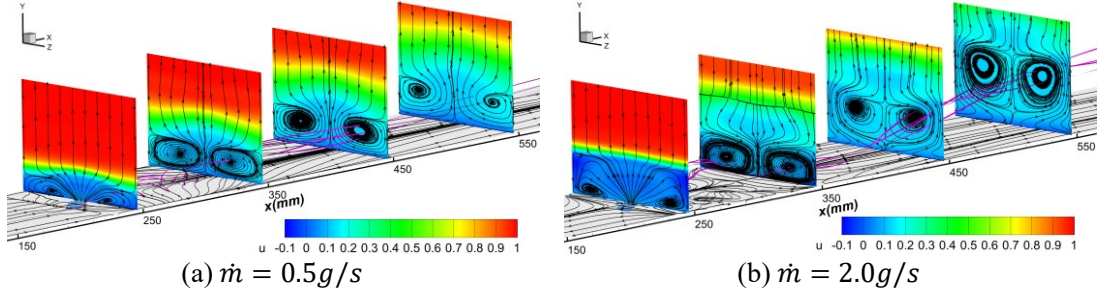

(a) $\dot{m} = 0.5 g/s$    (b) $\dot{m} = 2.0 g/s$

Fig. 8: Contours of $u$ and streamlines at the streamwise cross-sections (with $x = 250, 350, 450, 550 mm$) under different $\dot{m}$ and with $L_z = 15 mm$.

Fig. 9 presents the corresponding pressure contours and streamline distributions at the $z = 0$ cross-section and the wall for each of the cases. It can be observed that: (1) The injection induces a high-pressure region and shock-like structures above the hole, while an adverse pressure gradient is generated upstream of the hole, leading to flow separation and vortex formation. The resulting reverse flow contributes to friction reduction. Similar structures are also observed downstream of the injection region; (2) The injected flow forms a low-pressure region at the hole's exit, and there is a different degree of pressure increase downstream, followed by a downstream flow that exhibits a favorable pressure gradient; (3) Due to the symmetric boundary conditions in the spanwise direction, pressure increased near the spanwise boundary with the local maxima. The surface streamline patterns reveal mutual interactions between adjacent injections near the spanwise boundary; (4) As $\dot{m}$ increases, the high-pressure region (indicated in red) expands above and around the injection. The adverse pressure gradient strengthens, the separation region upstream of the injection enlarges, and the surface separation structures near the injection become more complex. For the case where $\dot{m} = 0.5 g/s$, the downstream spanwise boundary of the flat plate corresponds to a reattachment line, while the symmetry centerline ($z = 0$) represents a separation line. In contrast, for $\dot{m} = 2.0 g/s$, the separation lines shift to approximately $\pm L_z/2$ in the spanwise direction.

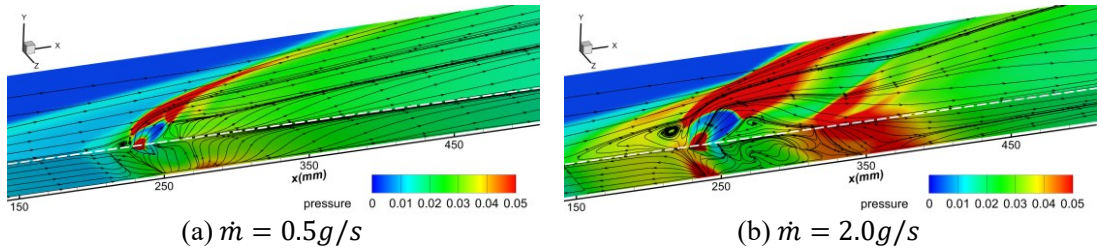

(a) $\dot{m} = 0.5 g/s$    (b) $\dot{m} = 2.0 g/s$

Fig. 9: Isobar contours and streamlines at the $z = 0$ cross-section and on the wall under different $\dot{m}$ and with $L_z = 15 mm$.

### 3.2.2 Patterns under different inter-hole spacing

To investigate flow structures under different hole spacings, two representative cases are analyzed, where $L_z = 15$ and $45 mm$ under a fixed $\dot{m} = 1.5 g/s$. Fig. 10 presents the contours of $u$ and streamline distributions at several streamwise cross-sections. The results show that: (1) In all cross-sections, the streamlines reveal the presence of a pair of dominant vortices extending over a relatively long-lasting streamwise region, with their vortex cores intersected by the same 3-D streamlines. Additional secondary vortex pairs appear in the vicinity of the main vortices at the $x = 350mm$ cross-section. A relatively greater number of vortices with increased complexities are observed in the case where $L_z = 15mm$; however, they decay rapidly along the streamwise direction; (2) Due to the imposed spanwise boundary conditions, the streamlines at the cross-sections generally remain in parallel to the spanwise boundary when approaching it; (3) For the case where $L_z = 15mm$, the boundary between high- and low-speed flows is relatively uniform and horizontally aligned. The $u$ exhibits an even distribution in the spanwise direction, with a low-speed layer near the wall. In contrast, the $L_z = 45mm$ case demonstrates a prominent bulge distribution of $u$ in the center, clearly indicating a spanwise non-uniformity, particularly in the near-wall region with low-speed flow; (4) In the case where $L_z = 15mm$, the vertical movement (or "lifting") of the vortex cores downstream of the injection is relatively smaller. For the case where $L_z = 45mm$, the dominant vortex pair in the $z > 0$ region rotates clockwise when viewed along the positive $x$ axis, which contrasts with that observed in the $L_z = 15mm$ case. Furthermore, the vortex cores are notably lifted away from the wall and exhibit strong induced motion. These characteristics suggest that variations in hole spacing $L_z$ may significantly influence the friction reduction of the air film.

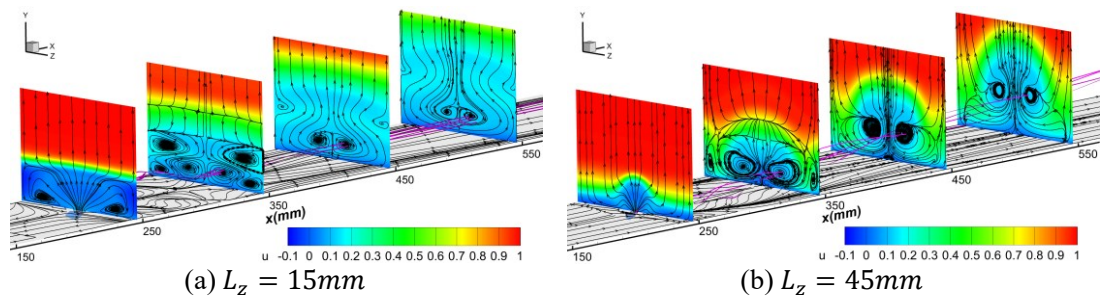

(a) $L_z = 15mm$          (b) $L_z = 45mm$

Fig. 10: Contours of $u$ and streamlines at the streamwise cross-sections (with $x = 250, 350, 450, 550mm$) under different $L_z$ and with $\dot{m} = 1.5g/s$.

Fig. 11 presents the corresponding isobar contours and streamline distributions at the $z = 0$ cross-section and the wall surface for each of the cases. It can be observed that, as in Fig. 9, the flow exhibits fundamental structures such as injection-induced shock waves and separation caused by adverse pressure gradients. Additionally, the downstream adverse pressure results in the flow separation displayed in the $z = 0$ cross-section, where a flow channel in the section is formed between two separation lines. Meanwhile, (1) In the $L_z = 15mm$ case, a larger separation region is observed upstream of the injection, while the downstream separation is relatively weaker. In contrast, the $L_z = 45mm$ case exhibits the opposite trend. In the latter scenario, the streamlines further converge downstream of the separated region behind the injection, thereby narrowing the injected flow channel in the cross-section; (2) In the $L_z = 15mm$ case, a stronger adverse pressure is generated near and downstream of the injection flank. The reattachment line is distributed along the spanwise direction, while reattachment nodes with large scale and new separation lines occur on the wall downstream of the injection. The adverse pressure is comparatively weaker in the $L_z = 45mm$ case. The reattachment line upstream of the injection forms a horseshoe-shaped pattern, while downstream separation lines converge near $z = 0$. While there exist separation-related saddle nodes, their scale is relatively small.

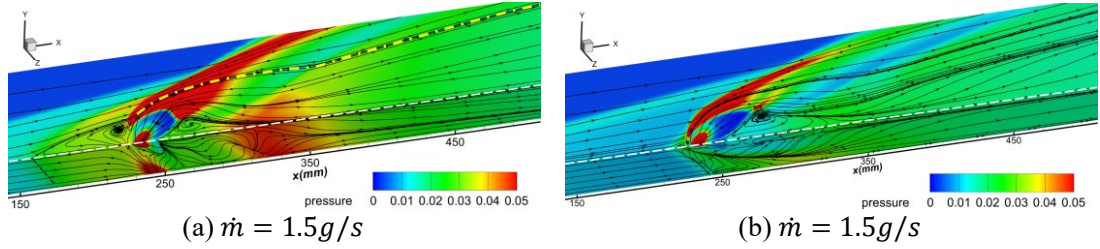

(a) $\dot{m} = 1.5 g/s$  (b) $\dot{m} = 1.5 g/s$

Fig. 11: Isobar contours and streamlines at the $z = 0$ cross-section and on the wall under different $L_z$ and with $L_z = 15mm$.

This subsection provides a comparative analysis of the flow structures associated with parallel injection under diverse $\dot{m}$ and $L_z$. This study reveals that increasing $\dot{m}$ expands the low $u$ layer above the wall surface, which is beneficial for enhancing friction reduction. Conversely, as the $L_z$ increases, the low $u$ layer become smaller and the spanwise non-uniformity becomes more pronounced, indicating that the air film has a negative effect on friction reduction.

### 3.3 Friction reduction performances under different mass flow rates and hole spacings

As demonstrated above, there are significant differences in flow field structures under diverse $\dot{m}$ and $L_z$. To quantitatively assess the friction reduction performance of the air film under various conditions, performances such as drag reduction efficiency and differences in $C_f$ between film and non-film cases are evaluated.

#### 3.3.1 Effect of mass flow rate

The drag reduction efficiency $\eta_D$ is first evaluated and compared under different $\dot{m}$ for the case where $L_z = 15mm$, which is defined as:

$$\eta_D = \frac{\iint C_{f,nojet} ds - \iint C_{f,jet} ds}{\iint C_{f,nojet} ds}, \tag{6}$$

where the subscripts "nojet" and "jet" refer to cases without and with injection, respectively. For a more efficient evaluation of the effect of the air film, the integration domain is defined from $100mm$ downstream of the plate leading edge to the trailing edge. Table 2 summarizes $\eta_D$ for various $\dot{m}$. Within the evaluation region, the air film significantly reduces wall shear stress, and $\eta_D$ exhibits an increasing trend with rising $\dot{m}$, reaching up to 66.17%. Specifically, $\eta_D$ increases markedly when $\dot{m} \leq 1.5 g/s$. However, when $\dot{m} = 2.0 g/s$, the improvement becomes relatively marginal.

Table 2: Drag reduction efficiency under different mass flow rates.

| $\dot{m}(g/s)$ | 0.5 | 1.0 | 1.5 | 2.0 |
|---|---|---|---|---|
| $\eta_D(\%)$ | 36.62 | 47.02 | 63.91 | 66.17 |

To illustrate the spatial distribution of friction reduction, the difference in $C_f$ is computed as $\Delta C_f = C_{f,jet} - C_{f,nojet}$. Fig. 12 presents the contours of $\Delta C_f$ and the corresponding streamline distributions for various $\dot{m}$, with the red isoline indicating $\Delta C_f = 0$. Given the symmetry of the problem with respect to the $z = 0$ plane, only half of the domain is shown to save space (a similar treatment is applied in Fig. 16 as well as in Sections 4 and 5). It can be observed that (1) The most significant friction reduction occurs upstream of the injection holes; in the far downstream field, the reduction effort is comparable across all cases and exhibits uniform distribution; (2) As $\dot{m}$ increases, both the magnitude and extent of the friction reduction upstream of the injections increase. Downstream, the region in which $C_f$ increases (in red) first diminishes and later reappears, shifting from the spanwise boundaries toward the centerline. This trend provides a qualitative indication of a weakening of the friction reduction enhancement with further increases in $\dot{m}$, which is consistent with the results in Table 2; (3) With increasing $\dot{m}$, the post injection flow structure becomes more complex; in the far field downstream, the separation lines shift from the wall centerline toward the spanwise boundaries.

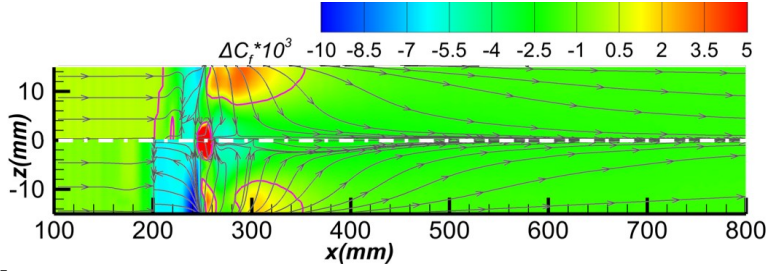

(a) $z \in [0,15]$ for the $\dot{m} = 0.5 g/s$ condition, while $z \in [-15,0]$ for the $\dot{m} = 1.0 g/s$ condition

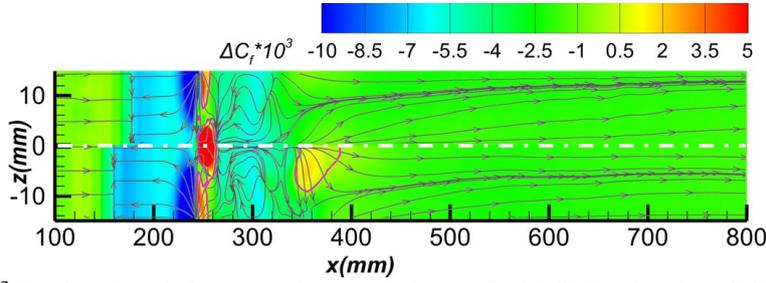

(b) $z \in [0,15]$ for the $\dot{m} = 1.5 g/s$ condition, while $z \in [-15,0]$ for the $\dot{m} = 2.0 g/s$ condition

Fig. 12: Contours of $\Delta C_f$ and streamlines under different $\dot{m}$ on the wall and with $L_z = 15mm$ (dashed lines working as the boundary to divide different cases).

To further quantify and compare the friction reduction performance at different streamwise locations, Fig. 13 presents the distribution of the spanwise averaged friction coefficient $\bar{C}_f$ with $x$ for various injections. $\bar{C}_f$ is defined as:

$$\bar{C}_f(x) = \frac{1}{2L_z} \int_{-L_z}^{L_z} C_f(x,z) \, dz. \tag{7}$$

In the figure, "NO-JET" denotes the same flat plate under the same inflow but without injection. The results indicate that: (1) Near the leading edge of the flat plate, $\bar{C}_f$ is consistent across all cases, suggesting that the air film has negligible influence in this region. A significant drop in friction is observed near the injection holes, with negative $\bar{C}_f$ occurring in some cases. A sharp peak of $\bar{C}_f$ appears at the injection location, with values exceeding those of the "NO-JET" case. In the downstream far field, all injection cases exhibit similarly smooth distributions with lower magnitudes compared to that of the "NO-JET"; (2) As $\dot{m}$ increases, the range and magnitude of friction reduction around the injection holes become more pronounced, and the peak $\bar{C}_f$ at the injection location increases accordingly.

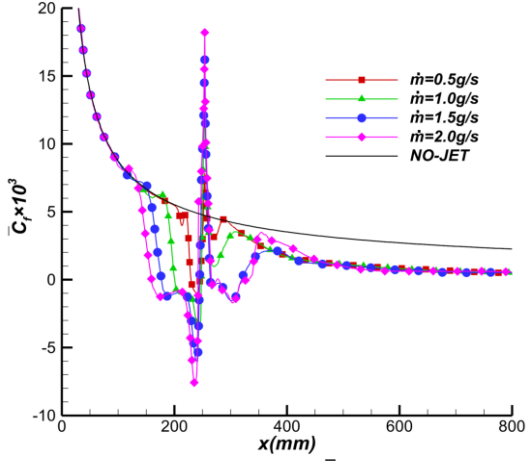 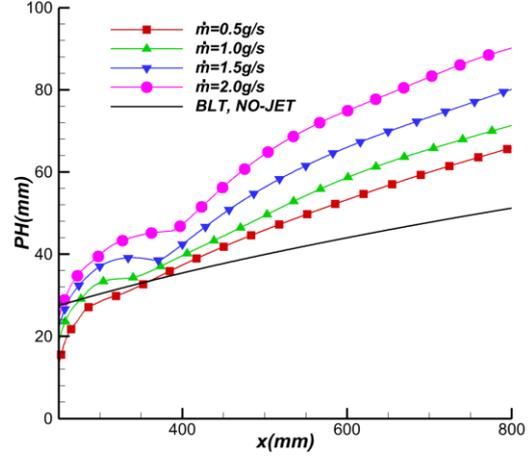

Fig. 13: Distributions of $\bar{C}_f$ with $x$ under different $\dot{m}$ and with $L_z = 15mm$.

Fig. 14: Distributions of PH under different $\dot{m}$ and with $L_z = 15mm$ as well as BLT with $x$.

Finally, Fig. 14 presents the streamwise distributions of penetration height (PH) for various $\dot{m}$, along with the boundary layer thickness (BLT) for the "NO-JET" case. BLT is measured where the local $u$ reaches 99% of $U_\infty$, i.e., $u/U_\infty = 0.99$, while the PH is determined based on the boundary dividing the injection and freestream streamlines on the $z = 0$ plane, as shown by the yellow dashed line in Fig. 11(a). For all injection flows, the PH exceeds the BLT of the "NO-JET" and increases with $x$. As $\dot{m}$ increases, the PH rises accordingly, indicating a greater "lift" of the freestream, which suggests a corresponding enhancement of friction reduction.

To quantify the flow losses induced by parallel injections, the total pressure ($P_0$) recovery coefficient along the streamwise cross-sections is computed. Given that $P_0$ usually varies across a cross-section in hypersonic flat plate flows, the mass-weighted $\bar{P}_0(x)$ serves as the evaluation metric, which is derived as follows:

$$\bar{P}_0(x) = \frac{\iint_\Omega \rho u P_0 ds}{\iint_\Omega \rho u ds}, \tag{8}$$

where $\Omega$ denotes the streamwise cross-sectional area of interest. The total pressure recovery coefficient is defined as $\sigma(x) = \bar{P}_0(x)/P_{0,\infty}$. Fig. 15 presents the streamwise distributions of $\sigma$ for various $\dot{m}$ with $L_z = 15mm$. To save space, the figure also presents the results for the cases where $L_z = 45mm$ and $\dot{m} = 1.5g/s$. It can be observed that all cases exhibit similar behavior upstream of the injections. Due to the injection of the air film, $\sigma$ decreases abruptly at the hole location, and the rate of decrease relaxes when it reaches approximately $x = 500$ mm. The overall decrease is relatively slow in the "NO-JET" case. Furthermore, an increase in $\dot{m}$ results in a lower $\sigma$, indicating greater flow losses.

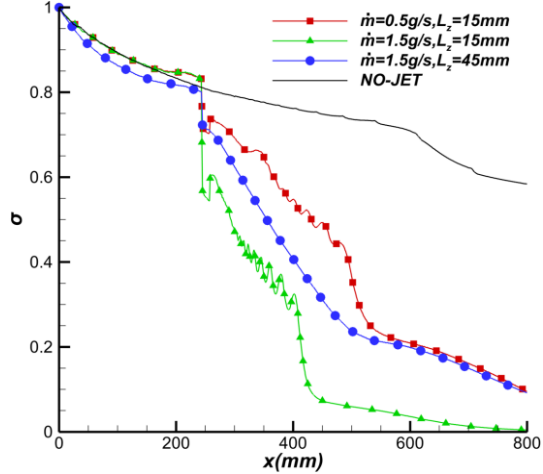

Fig. 15: Distributions of $\sigma$ with $x$ from different cases.

### 3.3.2 Effect of parallel hole spacing

Table 3 summarizes the $\eta_D$ for different $L_z$ under the condition $\dot{m} = 1.5g/s$, all of which demonstrate a pronounced reduction effect. According to the current investigation, $\eta_D$ decreases with increasing $L_z$. Specifically, the highest efficiency of 63.91% is achieved at the minimum spacing of $L_z = 15mm$, while this is nearly halved in the $L_z = 25mm$ case. As $L_z$ increases to 35 and 45 mm, $\eta_D$ still declines, albeit at a reduced rate.

Table 3: Drag reduction efficiency under different hole spacings.

| $L_z(mm)$ | 15 | 25 | 35 | 45 |
|---|---|---|---|---|
| $\eta_D(\%)$ | 63.91 | 33.07 | 22.05 | 20.51 |

Fig. 16 presents $\Delta C_f$ contour plots, as well as streamlines for various $L_z$, where the red isoline indicates $\Delta C_f = 0$. The results demonstrate that: (1) All configurations exhibit similar $\Delta C_f$ distributions, with prominent separation and recirculation regions near the injection holes where $\Delta C_f$ is significantly reduced, indicating obvious local friction reduction. Simultaneously, regions with increasing $C_f$ are also observed across all cases; (2) As $L_z$ increases, the friction reduction region in front of the hole gradually contracts and weakens, while the downstream regions adjacent to the hole experience more pronounced friction increase. In the region further downstream, the friction reduction appears to be relatively homogenic with a mild magnitude; (3) With increasing $L_z$, horseshoe-shaped reattachment lines emerge around the injection hole, spatially coinciding with the region of increased friction.

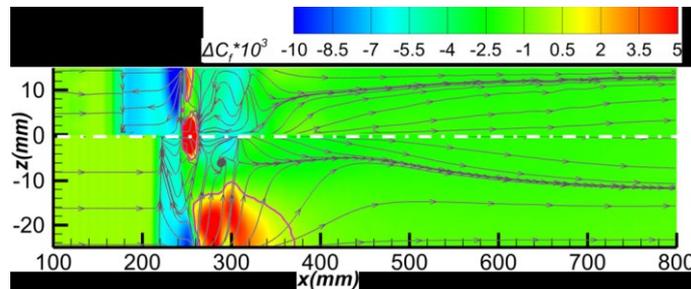

(a) $z \in [0,15]$ for the $L_z = 15mm$ case, while $z \in [-25,0]$ for the $L_z = 25mm$ case

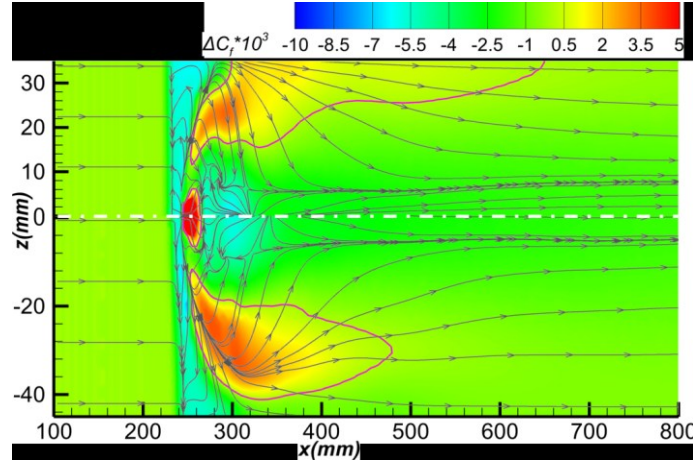

(b) $z \in [0,35]$ for the $L_z = 35mm$ case, while $z \in [-45,0]$ for the $L_z = 45mm$ case

Fig. 16: Contours of $\Delta C_f$ and streamlines under different $L_z$ on the wall and with $\dot{m} = 1.5g/s$ (dashed lines representing the boundary dividing different cases).

The streamwise distributions of $\bar{C}_f$ for various cases of $L_z$ are illustrated in Fig. 17. It can be observed that: (1) The results near the leading edge remain highly consistent across all cases. A pronounced drop and abrupt changes occur near the injection holes, with the peak located at the hole position. Downstream of the hole, $\bar{C}_f$ becomes smoother and quantitatively lower than that in the "NO-JET" case; (2) As $L_z$ increases, the magnitude and extent of the friction reduction upstream of the holes weakens, and the peak friction at the injection location decreases. In the near field downstream of the hole, substantial friction reduction is only observed for the $L_z = 15mm$ case, while the other configurations have a similar performance to one another. In the far downstream region, the reduction amplitude of $\bar{C}_f$ decreases, with the $L_z = 35,45mm$ cases exhibiting nearly identical distributions.

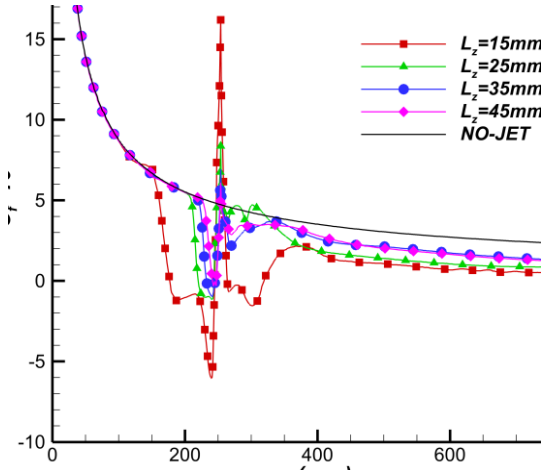

Fig. 17: Distributions of $\bar{C}_f$ with $x$ under different $L_z$ and with $\dot{m} = 1.5g/s$.

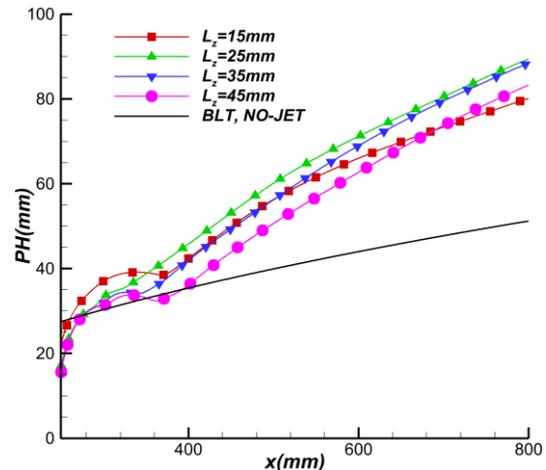

Fig. 18: Distributions of PH under different $L_z$ and with $\dot{m} = 1.5g/s$ as well as BLT with $x$.

Fig. 18 illustrates the distributions of PH for different $L_z$ conditions as well as the BLT of the case of no film with $x$, where overall growing distributions of PH are represented by values larger than BLT. Except for the $L_z = 15mm$ case, the PH generally decreases as $L_z$ increases. Although the case where $L_z = 15mm$ exhibits a relatively high PH near the injection, it subsequently transitions to a lower magnitude and slower rate of increase beyond approximately $x = 400mm$. Additionally, increasing $L_z$ leads to an improved $\sigma$ and thereby reduced flow losses as shown in Fig.

15.

To summarize, the friction reduction under different $\dot{m}$ and $L_z$ is quantitatively analyzed in this section. For variant $\dot{m}$, an increase in $\dot{m}$ leads to an expansion and enhancement of the friction reduction region near the holes, resulting in improved $\eta$. However, increased friction along the centerline limits further improvement. As the $L_z$ increases, a larger region with increased friction appears in the near field downstream of the holes, while the friction reduction in the far field decreases, causing an overall decline in friction reduction. When selecting injection parameters, it is necessary to balance the friction reduction requirement and acceptable levels of flow loss.

### 3.4 Mechanisms of friction reduction by air film in parallel injection

The friction reductions caused by air film differ quantitatively among different cases, and the underlying mechanisms will be analyzed in this section. In addition to the overall friction reduction, the discussion centers on the localized friction increase caused by the air film.

(1) Mechanisms of friction reduction

Fig. 19 presents contours of $u$ at the $z = 0$ and $z = -L_z$ sections for selected cases, as well as those of $\partial u/\partial y$ at the $z = -L_z/2$ section. The flow structures are observed to be similar, with the flow separation occurring upstream of the injection due to the film, followed by a reduction in velocity within the influence zone of the air film, and regions of high velocity gradient shifting away from the wall. Comparing Fig. 19(a) and (b), the following trends emerge with increasing $\dot{m}$: (1) The regions of low $u$ (depicted in blue) near the injection holes expand, and the thickness of this region downstream increases; (2) The thickness of the region of low $u$ at the $z = -15mm$ section increases, indicating an extended spanwise influence of the injection; (3) The transition zone (shown in green) between the region of low $u$ and the freestream narrows and shifts upward, suggesting that the stronger-shear zones are displaced away from the wall. These behaviors enhance the friction reduction effect of the air film. For cases with different $L_z$, a comparison between Fig. 19(d) and (c) reveals that: (1) In Fig. 19(d), the region of low $u$ upstream of the injection at the $z = 0$ section is smaller than that of Fig.19(c); (2) The region of low $u$ at the $z = L_z$ section of the former is narrower, indicating a reduced spanwise coverage of the air film. These outcomes adversely affect the friction reduction.

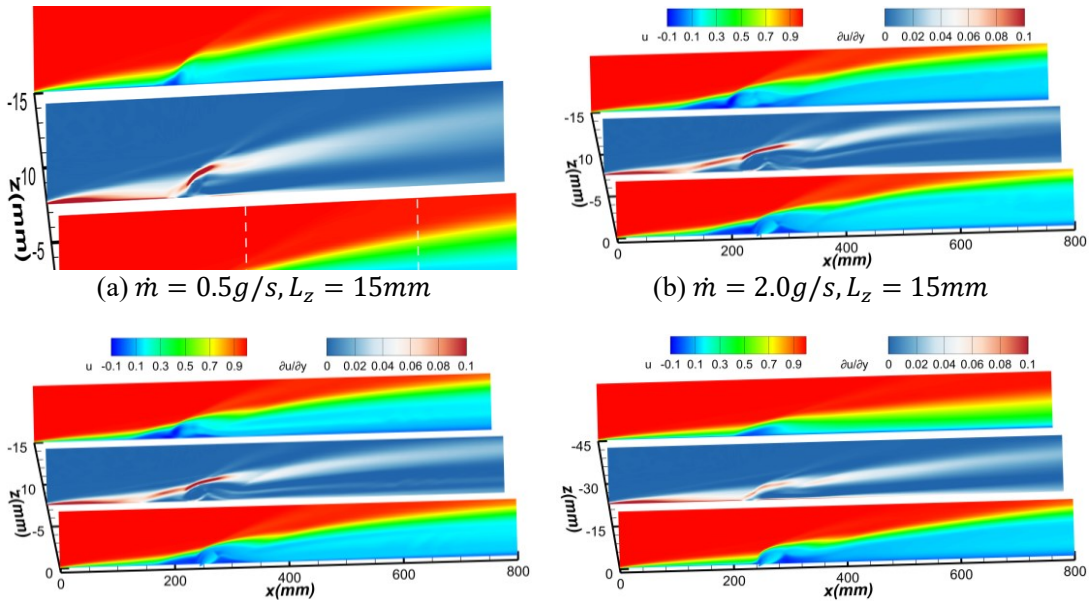

(a) $\dot{m} = 0.5g/s, L_z = 15mm$  (b) $\dot{m} = 2.0g/s, L_z = 15mm$

(c) $\dot{m} = 1.5g/s, L_z = 15mm$  (d) $\dot{m} = 1.5g/s, L_z = 45mm$

Fig. 19: Contours of $u$ at the $z = 0$ and $z = -L_z$ cross-sections, and those of $\partial u/\partial y$ at the $z = -L_z/2$ cross-section under different $\dot{m}$ and $L_z$.

To further analyze the mechanisms of friction reduction caused by air film, Fig. 20 illustrates the wall-normal distributions of $u$ at several center locations with $z = 0$ for the cases under investigation (the locations are indicated by the white dashed lines depicted in Fig. 19(a)). In the figure, the horizontal line labeled "PH" denotes the local PH for the respective cases. The results reveal a rapid increase of $u$ in the $y$ direction near the wall, where a plateau forms with $u/U_\infty = 0.1$, which is significantly lower than $U_\infty$. This is followed by a sharp acceleration approaching $U_\infty$. Across different locations, the extent of this plateau increases in the $x$ direction for the given condition, and the local PH consistently coincides with regions exhibiting significant $\partial u/\partial y$, exceeding the BLT of the "NO-JET" case. Notably, a novel shear layer with reduced velocity gradients emerges near the wall due to the air film injection, indicating a marked shear decrease compared to the "NO-JET" scenario, thereby contributing to friction reduction. Comparing Fig. 20(a) and (b), it can be observed that the $y$-coordinate of the position at which a given $u$ is attained increases with increasing $\dot{m}$, signifying a rise in the PH. Furthermore; at $x = 325mm$, negative $u$ appears near the wall for $\dot{m}$ of $1.5g/s$ and $2.0g/s$, indicating a local flow reversal that favors friction reduction somewhat. A comparison of Fig. 20(c) and (d) reveals that increasing the $L_z$ reduces the aforementioned $y$ coordinate, corresponding to a lower PH. It is noted that at the location $x = 625$ mm, the $L_z = 15mm$ case exhibits a notably thicker velocity transition to $U_\infty$, which is consistent with observations in Fig. 19(c) and (d).

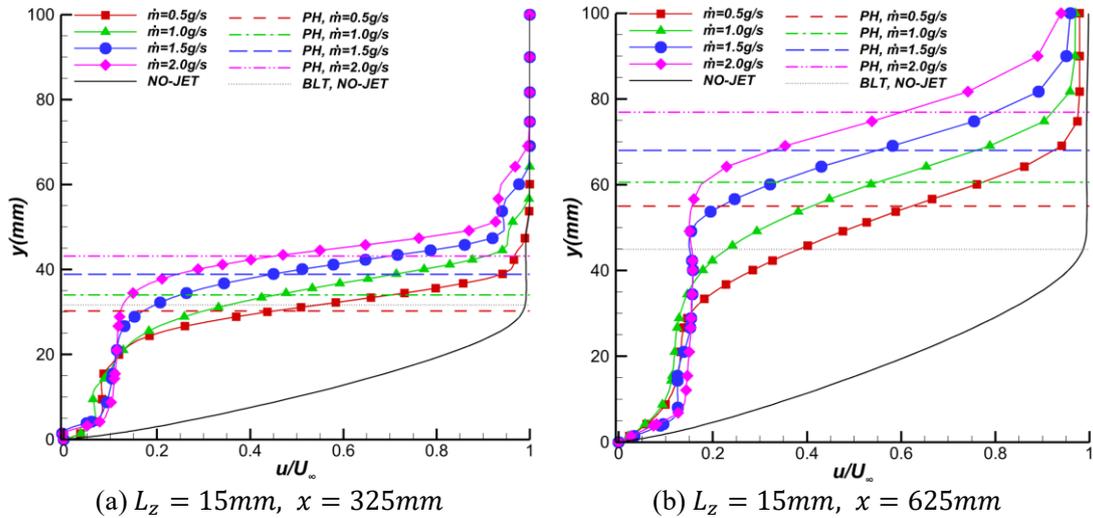

(a) $L_z = 15mm, \ x = 325mm$  (b) $L_z = 15mm, \ x = 625mm$

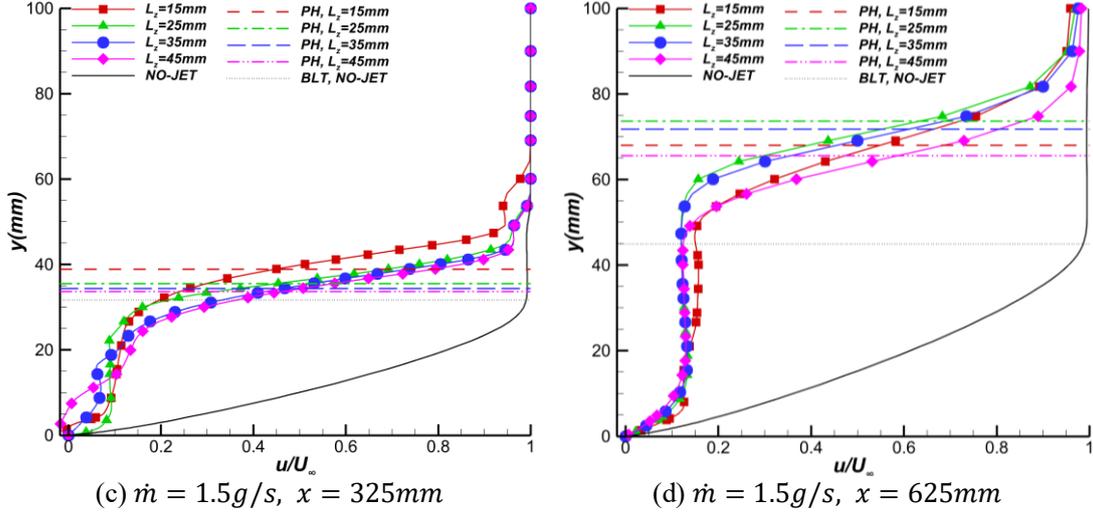

(c) $\dot{m} = 1.5 g/s$, $x = 325mm$  (d) $\dot{m} = 1.5 g/s$, $x = 625mm$

Fig. 20: Distributions of $u$ with $y$ at points with different $x$ at the $z = 0$ cross-section of various cases with corresponding PHs indicated by horizontal lines.

To investigate the distribution of $u$ at different spanwise locations, Fig. 21 takes the condition where $\dot{m} = 1.5 g/s, L_z = 45mm$ as representative and presents the wall-normal profiles of $u$ at various spanwise positions with $x = 625mm$. As the position approaches the lateral boundaries, the plateau feature in distributions gradually diminishes and the near wall velocity gradient progressively increases, yet remains lower than that in the "NO-JET" case.

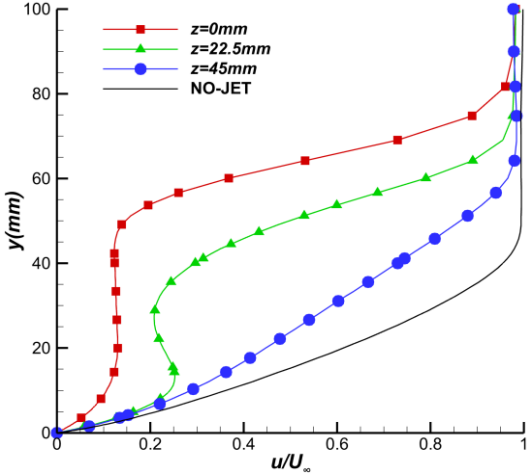 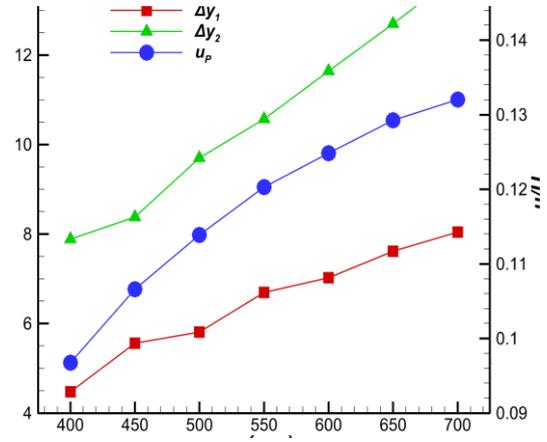

Fig. 21: Distributions of $u$ with $y$ at different spanwise locations with $x = 625mm$ of the $\dot{m} = 1.5 g/s$ and $L_z = 45mm$ case.

Fig. 22: Distributions of vortex layer thickness and $u_P$ with $x$ of the $\dot{m} = 1.5 g/s$ & $L_z = 45mm$ case.

To illustrate the evolution of the two shear layers, i.e., that near the wall and the other attaching the freestream as shown in Fig. 20, the streamwise distributions of their thickness are examined and the velocity of the $u$-plateau ($u_P$) is analyzed using the same case where $\dot{m} = 1.5 g/s$ and $L_z = 45mm$ case. The thickness is defined as that of the vortex layer:

$$\Delta y = \Delta u / max(\partial u / \partial y), \tag{9}$$

where $max(\cdot)$ runs through the layer and $\Delta u$ represents the velocity difference across it. Specifically, $\Delta y_1$ corresponds to the layer near the wall and is derived by taking $\Delta u = u_P - 0$, and $\Delta y_2$ represents the other layer and is computed using $\Delta u = (U_\infty - u_P)/2$. Fig. 22 presents the corresponding distributions, which show an increasing trend of both thicknesses along the $x$-direction, with $\Delta y_1$ being approximately half of $\Delta y_2$.

In summary, the mechanism investigation reveals that the injection of air film engenders a plateau in distributions of $u$ with $y$ near the wall, which has a value significantly lower than $U_\infty$. This leads to a reduction of $\partial u/\partial y$ and induces the friction reduction. As $\dot{m}$ increases, the extent of the near-wall region of low $u$ expands, and the height/scale of the velocity plateau increases; with increasing $L_z$, this near wall region narrows, leading to a weakened friction reduction.

(2) Discussions about local friction increase often neglected by investigations

As noted in Section 3.3, all air films generated by injections exhibit local increases in friction, which are commonly located near or along the reattachment of separation. The case where $\dot{m} = 1.5g/s$ and $L_z = 45mm$ is selected and the subsequent analysis is illustrated in Fig. 23. To improve the discussion, the lower half of Fig. 16(b) is copied to comprise the lower half of Fig. 23 ($z \in [-45,0]$), while the upper half ($z \in [0,45]$) presents the $C_f$ contours with streamlines still depicted. In the figure, an isoline of $C_f = 0.0031$ in the region $z > 0$ corresponds approximately to the isoline $\Delta C_f = C_{f,jet} - C_{f,nojet} = 0$ at $z < 0$, and this isoline encompasses the bulk of the primary reattachment. It is therefore inferred that the shear of $u$ is intensified during the reattachment process, resulting in a local increase in $C_f$ compared to the "NO-JET" case.

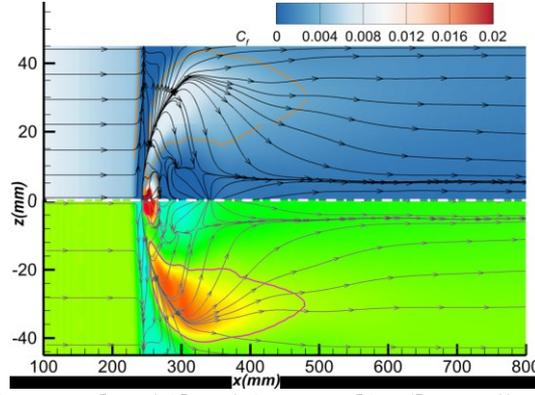

Fig. 23: Contours of $\Delta C_f$ at $z \in [-45,0]$ and $C_f$ at $z \in [0,45]$ as well as streamlines on the wall.

## 4 Flow structures and friction reduction in tandem injections under various conditions

Section 3 examined the effects of parameters, such as $\dot{m}$ and $L_z$, on friction reduction in parallel injections. Similar issues are also relevant to air film generated by tandem injections. In this section, the JET2 (see Section 3.1.1) is again adopted as the object for investigation, and simulations are conducted for tandem injections with diverse $\dot{m}$ and $L_h$. The effects on flow structures and friction reduction are analyzed, and the underlying mechanisms are discussed.

### 4.1 Setup of tandem injections

Fig. 24 illustrates the tandem injections used in this study, which involve three injection holes arranged in tandem along the centerline of the flat plate. The flat plate has the same length as that used in the parallel injections, with a spanwise width of $L_z = 50mm$. The first hole is located at a distance of $L_x = 250mm$ from the leading edge of the plate, and the streamwise spacing between tandem holes is denoted by $L_h$. The geometric parameters of the injection holes, including size and inclination angle, are identical to those of the JET2. Outflow boundary conditions are applied at both spanwise sides, while other settings follow those detailed in Section 3.1.1. The computational cases are established as follows: first, simulations are conducted at $L_h = 150mm$ with various $\dot{m}$ of $0.1, 0.2, 0.3, 0.4 g/s$. Subsequently, at a fixed $\dot{m}$ of $0.3 g/s$, simulations are performed for different $L_h$s of $150, 175, 200mm$. It is worth noting that $\dot{m}$ used in the tandem injections are lower than those in the parallel configuration. This is primarily due to two considerations: first, to ensure that the total $\dot{m}$ of all injections on the flat plate is comparable to those of parallel injections—a greater number of injections in the tandem setup necessitates a reduction of the $\dot{m}$ of individual injections; second, a larger $\dot{m}$ would significantly intensify the interference between injections, aggravating numerical stability. Computational practice indicates that the chosen $\dot{m}$ will ensure stable and reliable results.

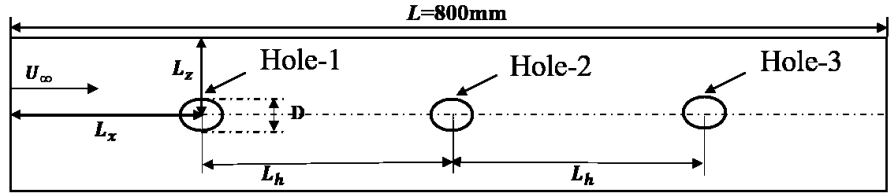

Fig. 24: Tandem injections and hole arrangement.

**4.2 Flow structures under different conditions**

**4.2.1 Patterns under different mass flow rates**

The cases where $\dot{m} = 0.1,\ 0.4 g/s$ are selected as representatives under the condition $L_h = 150mm$ to analyze the flow structures under tandem injections with different $\dot{m}$. Fig. 25 presents the contours of $u$ and streamlines at cross-sections located at each injection hole and at a distance of $L_h/2$ downstream of each hole. The results show that, as the flow develops downstream, the low $u$ layer exhibits a bulged profile with a pronounced central elevation, the extent of which increases streamwise as the near wall region with low $u$ expands. Due to the iterative feature of the tandem injections, similar streamline patterns are observed for each pair of cross-sections at the injection hole and the subsequent downstream section. Specifically, the streamlines at the injection section exhibit a source-like half-node that travels downstream and generates a pair of symmetric vortices. The vortices will be disturbed when they further pass through the downstream hole, and this process repeats as a typical pattern. Furthermore, as they evolve downstream, the flow structures in the sections between holes appear as symmetric vortices through which the identical 3-D streamlines penetrate, suggesting correlations among the vortical structures despite the disturbance of the injection. With increasing $\dot{m}$: (1) The extent of the low $u$ near wall region expands, the central bulging becomes more pronounced, the transition region (in green) becomes thinner, and the vertical elevation of the incoming flow increases; (2) The vortical structures visualized by streamlines on sections downstream of each hole increase in size and experience greater uplift as the flow moves downward. The downstream injections thus have a greater effect on the vortices, resulting in more complex disturbances.

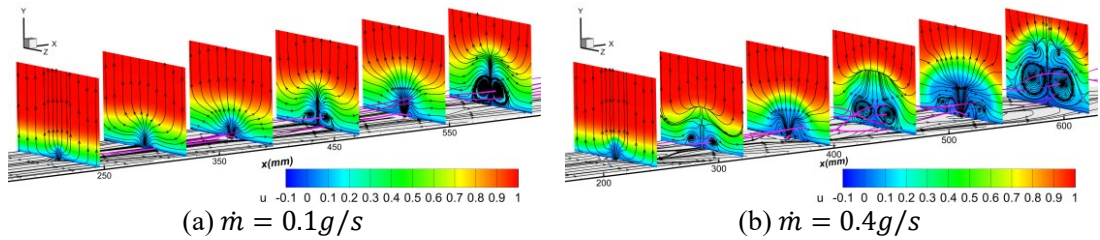

(a) $\dot{m} = 0.1 g/s$      (b) $\dot{m} = 0.4 g/s$

Fig. 25: Contours of $u$ and streamlines at the streamwise cross-sections (with $x = 250, 325, 400, 475, 550, 625 mm$) under different $\dot{m}$ and with $L_h = 150mm$.

Fig. 26 presents the corresponding pressure contours and streamlines on the spanwise plane at

$z = 0$ and on the wall surface. It can be observed that, because of the tandem injections, the flow structures around each hole are similar; that is, a distinct high-pressure region appears above each hole, while the low-pressure zone after the hole and the subsequent recirculation both grow in size as the flow moves downstream. On the wall surface, a clear separation line forms upstream of the first hole, followed by the horseshoe-shaped reattachment line arising and extending downstream. Meanwhile, with increasing $\dot{m}$: (1) The high-pressure regions upstream of each hole in the $z = 0$ section expand and intensify, the low pressure zones after the hole become larger, the downstream recirculation region grows, and the wall pressure increases overall; (2) The incoming flows are more elevated over the holes; on the wall surface, the downstream saddle points behind Holes-2 and -3 (denoted as "SP" in the figure) shift further downstream, while the upstream saddle points move forward. The reattachment line on the wall expands, indicating that the injection has a broader influence range.

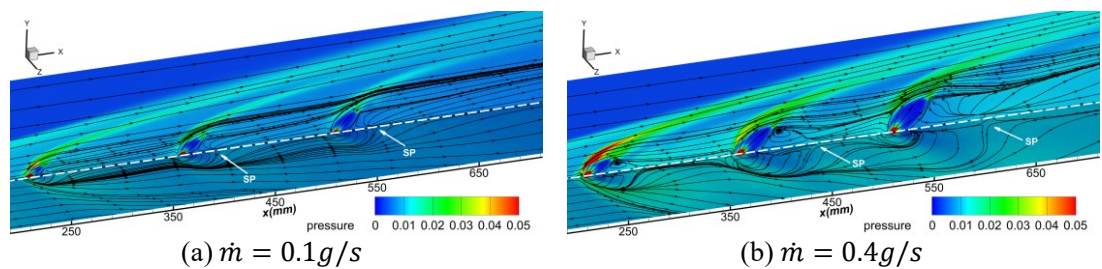

(a) $\dot{m} = 0.1 g/s$      (b) $\dot{m} = 0.4 g/s$

Fig. 26: Isobar contours pressure and streamlines at the $z = 0$ cross-section and on the wall under different $\dot{m}$ and with $L_h = 150mm$.

**4.2.2 Patterns under different inter-hole spacing**

To investigate the flow structures under different $L_h$, the cases where $L_h = 150, 200mm$ are selected as representative under the condition $\dot{m} = 0.3g/s$, as shown in Fig. 27. The figure presents the contours of $u$ and streamlines on streamwise cross-sections located similarly to those presented in Section 4.2.1. The flow field downstream of the holes also exhibits a bulged velocity distribution, with a region of low $u$ near the wall, which overall resembles that shown in Fig. 25(b). As the $L_h$ increases, the flow structures at the injection planes remain largely similar, while some differences emerge in the downstream sections behind the hole. Specifically, in the case where $L_h = 150mm$, the vortex downstream of Hole-1 appears to be larger in scale, and an additional saddle point and separation-type node are observed above the hole, while the $L_h = 200mm$ case exhibits more complex structures in the sections downstream of Holes-2 and -3.

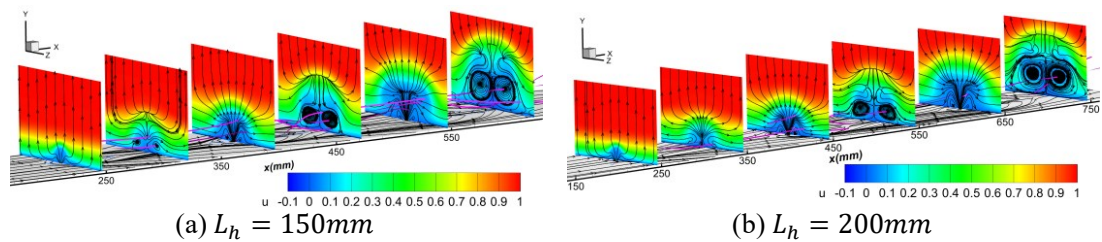

(a) $L_h = 150mm$      (b) $L_h = 200mm$

Fig. 27: Contours of $u$ and streamlines on the streamwise cross-sections under different $L_h$ and with $\dot{m} = 0.3g/s$.

Fig. 28 presents the corresponding pressure contours and streamlines in the $z = 0$ section and on the wall surface. The overall flow structures for different $L_h$ are similar to those observed in Fig. 26(b). As the $L_h$ increases, the size of the recirculation regions and reattachment lines on the wall near Holes-2 and -3 increases accordingly and, in contrast, the reattachment line near the centerline behind Hole-1 becomes less pronounced.

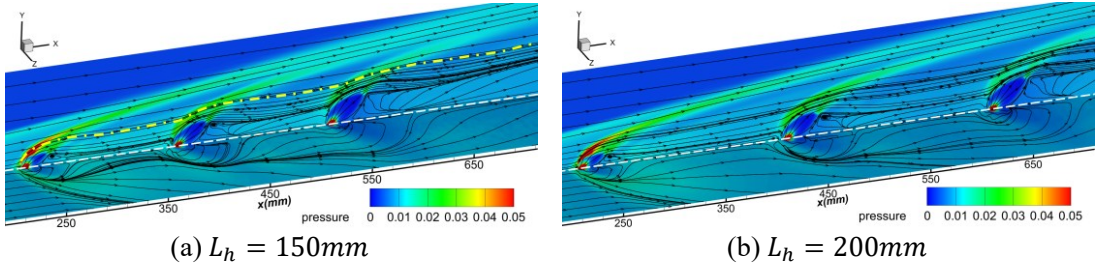

(a) $L_h = 150mm$  (b) $L_h = 200mm$

Fig. 28: Isobar contours and streamlines at the $z = 0$ cross-section and on the wall under different $L_h$ and with $\dot{m} = 0.3g/s$.

In summary, flow structures under different $\dot{m}$ and $L_h$ are analyzed in this section. The results indicate that as $\dot{m}$ increases, the lifting of the incoming flow becomes more pronounced, and the spanwise coverage of the air film expands. With increasing $L_h$, the differences in the overall flow structures remain relatively minor.

### 4.3 Friction reduction performances under different mass flow rate and hole spacing

As shown above, the flow structures clearly vary with diverse $\dot{m}$ but less obviously with different $L_h$. Their friction reduction performance is quantitatively analyzed in the following section.

#### 4.3.1 Effect of mass flow rate

The friction reduction performance, i.e., $\eta_D$ (see Eq. (6)) is first analyzed under different $\dot{m}$ and under the condition $L_h = 150mm$ as shown in Table 4. In terms of the effective coverage of the air film, the integration domain extends from $50mm$ upstream of Hole-1 to the trailing edge of the flat plate. All cases exhibit an obvious friction reduction, with the maximum $\eta_D$ reaching up to 38.62%. When $\dot{m} \leq 0.3g/s$, the $\eta_D$ increases with $\dot{m}$. However, when the $\dot{m}$ further increases to $\dot{m} = 0.4g/s$, the $\eta_D$ slightly decreases.

Table 4: Drag reduction efficiency under different mass flow rates.

| $\dot{m}(g/s)$ | 0.1 | 0.2 | 0.3 | 0.4 |
|---|---|---|---|---|
| $\eta_D(\%)$ | 23.97 | 32.97 | 38.62 | 36.14 |

To visualize the distribution of friction reduction, Fig. 29 presents the contours of $\Delta C_f (C_{f,jet} - C_{f,nojet})$ for the corresponding cases, with the red lines indicating $\Delta C_f = 0$. Significant friction reduction can be observed near the separation lines and downstream of each injection. However, the reduction in friction on the flank of each hole is relatively limited. Regions of increased friction appear downstream of the reattachment line behind Hole-1, as well as in some regions further downstream. Additionally, as $\dot{m}$ increases, the regions with friction reduction expand and the downstream zones with $\Delta C_f \geq 0$ after Hole-3 vanish, whereas such zones around both sides of Holes-1 and -2 become more pronounced.

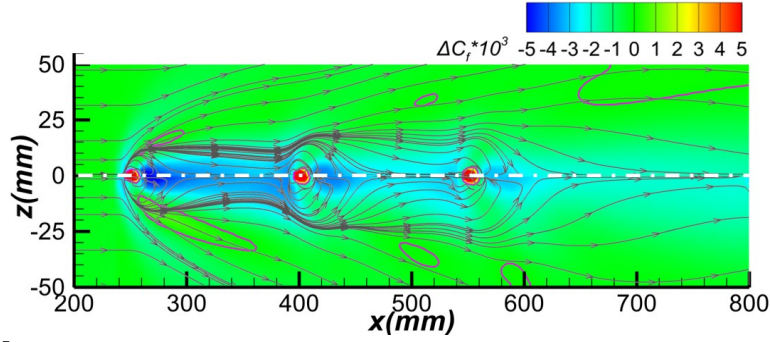

(a) $z \in [0,50]$ for the $\dot{m} = 0.1 g/s$ condition, while $z \in [-50,0]$ for the $\dot{m} = 0.2 g/s$ condition

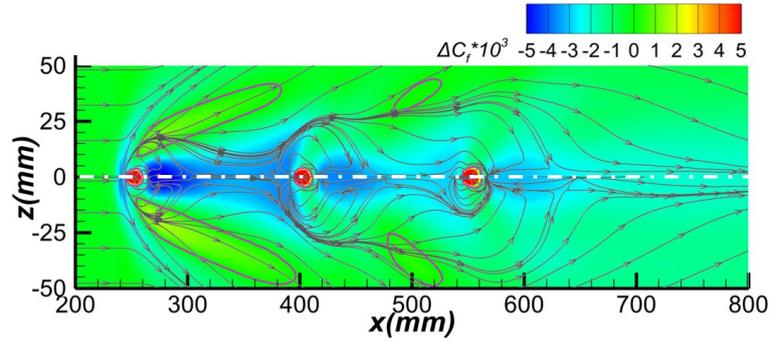

(b) $z \in [0,50]$ represents the $\dot{m} = 0.3 g/s$ condition, while $z \in [-50,0]$ represents the $\dot{m} = 0.4 g/s$ condition

Fig. 29: Contours of $\Delta C_f$ and streamlines under different $\dot{m}$ on the wall (dashed lines working as the boundary to divide different cases) and with $L_h = 150 mm$.

Fig. 30 presents the streamwise distributions of $\bar{C}_f$ (see Eq. (7)). The results indicate the following: (1) $\bar{C}_f$ generally decreases along the streamwise direction, while all distributions coincide near the leading edge of the flat plate, indicating minimal injection influence there. A sharp drop in $\bar{C}_f$ occurs just upstream of the injection holes, followed by a sudden increase; (2) With increasing $\dot{m}$, $\bar{C}_f$ exhibits an overall decreasing trend; however, all distributions upstream of Holes-1 and -2 remain relatively close.

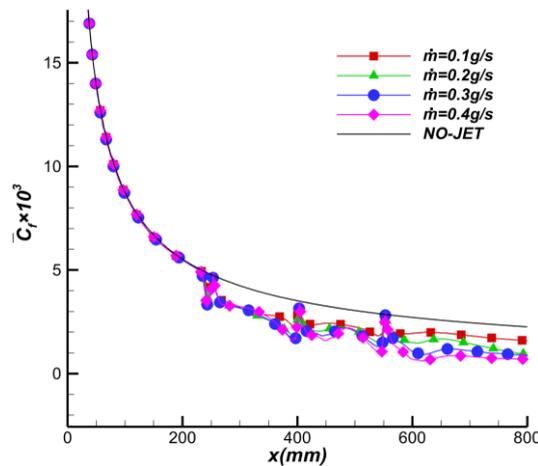

Fig. 30: Distributions of $\bar{C}_f$ with $x$ under different $\dot{m}$ and with $L_h = 150 mm$.

Fig. 31 presents the streamwise distributions of PH at $z = 0$ under different $\dot{m}$, along with the BLT for the "NO-JET" case, recalling that the PH is derived via the yellow dashed line as shown in Fig. 28(a). Both PH and BLT exhibit an overall increasing trend along the streamwise direction, and the larger $\dot{m}$ is, the larger the PH. Other than in the downstream regions of the cases where $\dot{m} = 0.3, 0.4 g/s$, the PHs are generally distributed below the BLT. When an injection is met, the PH rises sharply and subsequently transitions into a mild variation.

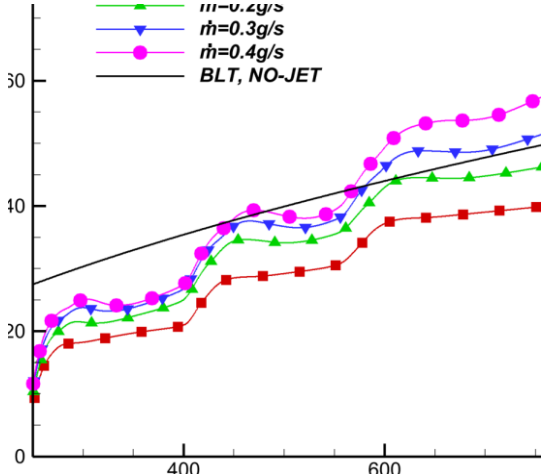 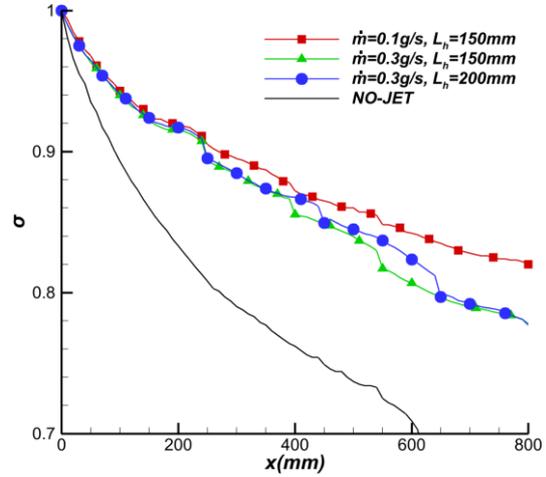

Fig. 31: Distributions of PH under different $\dot{m}$ and with $L_h = 150mm$ as well as BLT with $x$.

Fig. 32: Distributions of $\sigma$ with $x$ under different cases.

Fig. 32 presents the streamwise distributions of $\sigma$ to evaluate the flow losses in tandem injections under two $\dot{m}$, including the results with different $L_h$ to save space. All $\sigma$ exhibit a gradual streamwise decrease, with pressure loss relatively lower than that in the "NO-JET" configuration, implying an energizing by injections at small $\dot{m}$. As $\dot{m}$ increases, the $\sigma$ decreases accordingly after the injections, indicating larger flow losses.

**4.3.2 Effect of tandem hole spacing**

To investigate the effect of $L_h$, Table 5 presents $\eta_D$ for different $L_h$ under the condition $\dot{m} = 0.3 g/s$. Within the scope of the study, $\eta_D$ varies between 36.94% and 38.62%, exhibiting a slight decline as the $L_h$ increases, whereas there are small differences among cases.

Table 5: Drag reduction efficiency under different hole spacings.

| $L_h (mm)$ | 150 | 175 | 200 |
|---|---|---|---|
| $\eta_D (\%)$ | 38.62 | 37.31 | 36.94 |

Fig. 33 presents contours of $\Delta C_f$ for various cases, where the red isolines denote $\Delta C_f = 0$. The patterns of $\Delta C_f$ and streamlines resemble those shown in Fig. 29. As $L_h$ increases, there is an expansion of the blue regions indicating friction reduction between the holes, yet the areas of $\Delta C_f \geq 0$ on both sides also enlarge, resulting in an overall slight decrease in friction reduction.

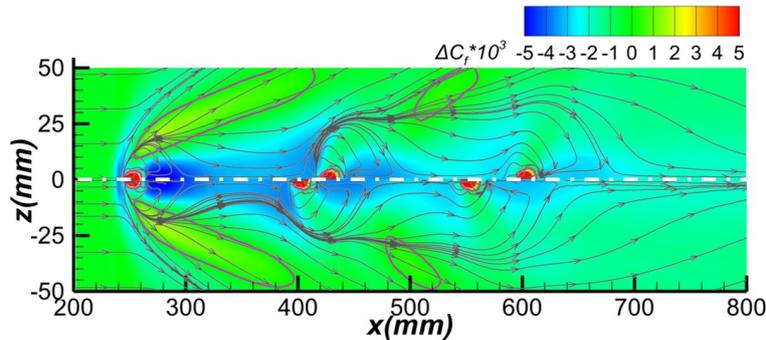

(a) $z \in [0,50]$ for the $L_h = 175mm$ case, while $z \in [-50,0]$ for the $L_h = 150mm$ case

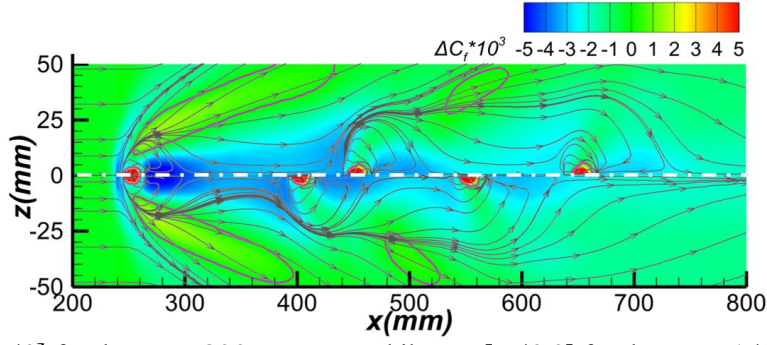

(b)$z \in [0,50]$ for the $L_h = 200mm$ case, while $z \in [-50,0]$ for the $L_h = 150mm$ case

Fig. 33: Contours of $\Delta C_f$ and streamlines under different $L_h$ on the wall and with $\dot{m} = 0.3g/s$ (dashed lines working as the boundary to divide different cases).

Fig. 34 presents the streamwise distributions of $\bar{C}_f$ for various $L_h$ cases, which are similar to those presented in Fig. 30, demonstrating that the air film generated by tandem injections induces friction reduction. The main difference lies in the location of abrupt changes, which shifts along with the hole positions.

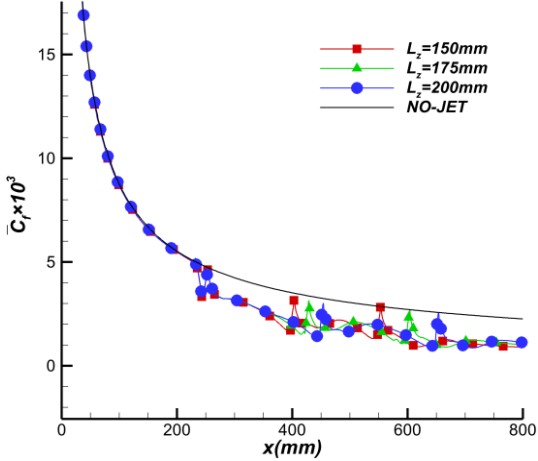

Fig. 34: Distributions of $\bar{C}_f$ with $x$ under different $L_h$ and with $\dot{m} = 0.3g/s$.

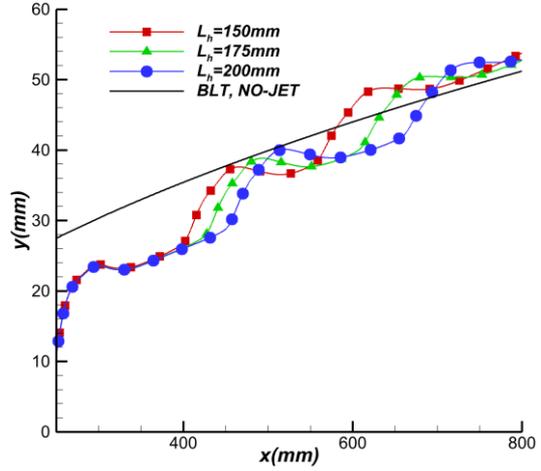

Fig. 35: Distributions of PH under different $L_h$ and $\dot{m} = 0.3g/s$ as well as BLT with $x$.

Similarly, Fig. 35 presents the streamwise distributions of PH for cases with different $L_h$, alongside the BLT for the case without a film. For all cases, the PH exhibits a stepwise increase along the streamwise direction, remaining generally lower than the BLT except in the downstream region. As shown in Fig. 32, as $L_h$ increases, the location of the rapid drop of $\sigma$ shifts downstream in accordance with the locations of the corresponding holes.

In summary, Section 4.3 quantitatively analyzes the friction reduction with different $\dot{m}$ and $L_h$, showing that as $\dot{m}$ increases, the friction reduction zone downstream of the injection enlarges, whereas the regions of $\Delta C_f \geq 0$ increase, and $\eta_D$ also initially increases accordingly, before a slight decline. With increasing $L_h$, the overall $\eta_D$ experiences only a marginal decrease. Given that there were no significant differences in flow losses among different $\dot{m}$ and that the variation in $\eta_D$ was minor across diverse $L_h$, the configuration where $\dot{m} = 0.3g/s$ and $L_h = 200mm$ is tentatively recommended.

### 4.4 Mechanisms of friction reduction by air film in tandem injection

Following the analysis in Section 3.4, Fig. 36 presents the contours of $u$ at the $z = 0$ and $z = $

$-L_z$ planes, as well as $\partial u/\partial y$ at the $z = -L_z/2$ section for various cases. It can be observed that, influenced by the tandem injections, the incoming flow exhibits pronounced lifting at each injection hole, followed by relatively smooth changes, until the next injection is met and a stepwise layout appears. Regions of low $u$ emerge near the wall downstream of the injection, effectively isolating the intense shear layer and distancing it from the wall, thereby contributing to friction reduction. Simultaneously, with increasing $\dot{m}$: (1) The lifting of the incoming flow by the air film increases, and the stronger shear region (depicted in red) at the $z = -25mm$ plane is located further from the wall; (2) The downstream region of low $u$ at the $z = -50mm$ plane expands, indicating an increased spanwise coverage of the air film. These occurrences thus enhance the friction reduction. Meanwhile, no obvious differences in friction reduction are observed among different $L_h$ cases. A comparison between Fig. 36(c) and (d) reveals similar velocity distributions and spanwise coverage of the air film, other than the different positions of the inflow lifting owing to the different $L_h$.

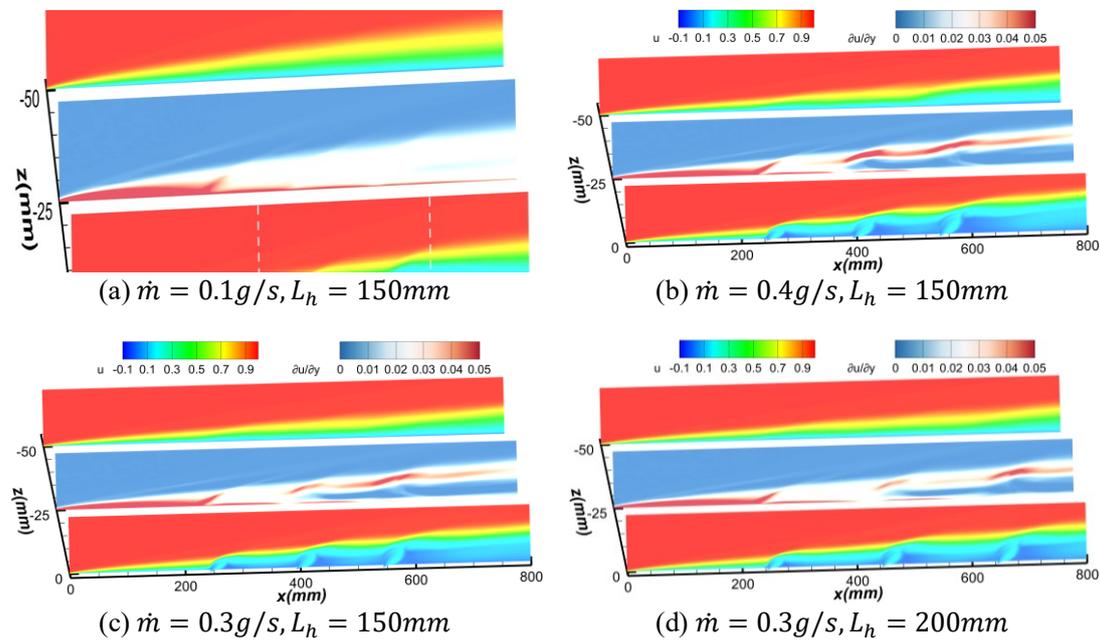

(a) $\dot{m} = 0.1g/s, L_h = 150mm$

(b) $\dot{m} = 0.4g/s, L_h = 150mm$

(c) $\dot{m} = 0.3g/s, L_h = 150mm$

(d) $\dot{m} = 0.3g/s, L_h = 200mm$

Fig. 36: Contours of $u$ at the $z = 0$ and $z = -50mm$ cross-sections, and those of $\partial u/\partial y$ at the $z = -25mm$ cross-section under different $\dot{m}$ and $L_h$.

To further investigate the differences in friction reduction performance under diverse $\dot{m}$, Fig. 37 illustrates the wall normal distributions of $u$ at selected center locations with $z = 0$ as indicated by the white dashed lines in Fig. 36(a). The results reveal that, similar to the parallel injections, for the two shear layers—the one near the wall and the other attached to the freestream—the $u$ of the former increases slowly while that of the latter increases rapidly. Due to the tandem injection configuration, no pronounced low-speed plateau is observed between the two layers, which is somewhat different from those shown in Figs. 19 and 20. The results further indicate that $\partial u/\partial y$ within the near wall shear layer is significantly reduced compared to that of the "NO-JET" case, resembling the occurrence with parallel injections. As $\dot{m}$ increases, the $y$ coordinate at which a given $u$ is attained also increases, and the PH rises. Considering the small differences in friction reduction among the various $L_h$, the distributions of $u$ with $y$ will be omitted accordingly.

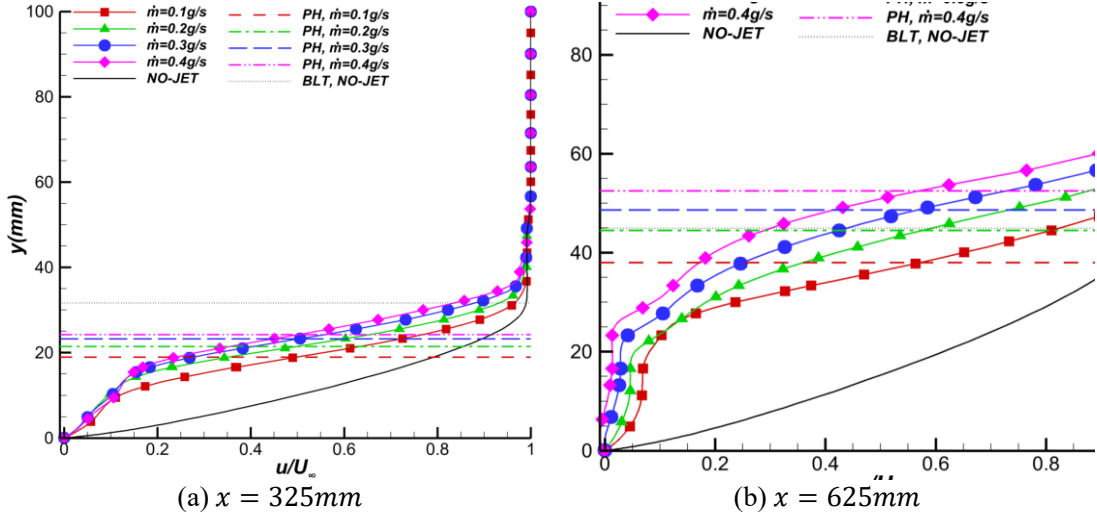

(a) $x = 325mm$  (b) $x = 625mm$

Fig. 37: Distributions of $u$ with $y$ at points with different $x$ on the $z = 0$ cross-section of various cases with corresponding PH indicated by horizontal lines.

To elucidate the spanwise variations of the normal distribution of $u$, Fig. 38 presents the distributions at several spanwise positions with $x = 750mm$, using the case where $\dot{m} = 0.3g/s$ and $L_h = 200mm$ as an example. As the location shifts toward the lateral boundary, the plateau feature of the velocity gradually diminishes, and noticeable fluctuations are observed in the distribution at $z = 25mm$. The near wall velocity gradient increases gradually when approaching the boundary, which is still lower than that of the "NO-JET" case.

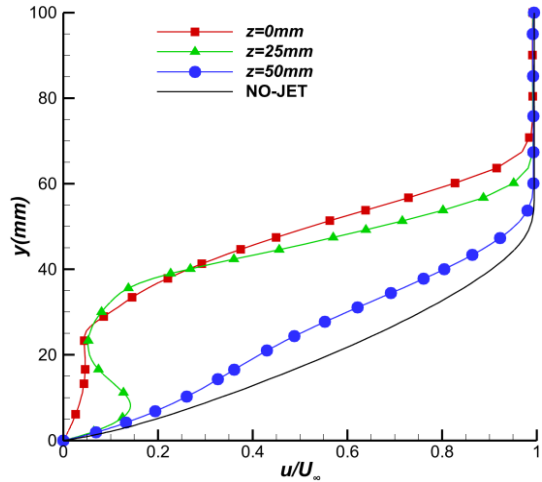

Fig. 38: Distributions of $u$ with $y$ at different spanwise locations at cross-section $x = 750mm$.

In addition to friction reduction, a local increase of friction force is also observed, as illustrated in Fig. 29. As analyzed in Section 3.4, this local increase is at least partially attributed to flow reattachment downstream of the injection.

## 5 Effect of hole's aspect ratio on friction reduction by parallel and tandem injections

As mentioned previously, the geometry of the injection holes has a significant impact on the friction reduction caused by air films, which is clearly valuable for engineering. In this study, the AR of an injection hole is targeted as representative of its geometric effects, which is accomplished by examining the diverse ARs of both parallel and tandem injection holes. The geometric details have been described in Section 3.1.1. Based on the insights obtained from Sections 3 and 4, the numerical conditions of the study are defined as follows: for parallel injection, $\dot{m} = 1.5 g/s, L_z = 45mm$, and for tandem injection, $\dot{m} = 0.3 g/s, L_h = 200mm$, with the other conditions remaining consistent with those in Section 3.1.1. As JET2 has been studied in detail in Sections 3 and 4, related content will not be repeated here. Furthermore, given the overall similarity in flow structures across different hole geometries, the differences owing to the AR of a hole are the primary concern of this section.

## 5.1 Effect of aspect ratio in parallel injection

(1) Effect on flow structures

To compare the flow structures from injections with different ARs, Fig. 39 presents the contours of $u$ and streamlines at four streamwise cross-sections, where a pair of counter-rotating primary vortices is featured. In the JET1 case, the primary vortices exhibit a larger spanwise extent. High-velocity fluid in the transition zones (green regions) is more entrained downward by the vortices toward the near-wall region, while the low-speed fluid between the two vortices is lifted upward, forming a pronounced bulge of moderate $u$. This distribution indicates significant spanwise non-uniformity in $u$ as well as the strong shear stress near the wall. In contrast, the JET3 case exhibits primary vortices with reduced spanwise size. At the $x = 350mm$ cross-section, an additional pair of vortices with counter rotation to the primary vortices emerges near the lateral boundaries. These secondary vortices weaken the entrainment of the transitional flow by the primary vortices, resulting in improved spanwise uniformity of $u$ and an enlarged near-wall low-speed region, both of which are likely to enhance friction reduction.

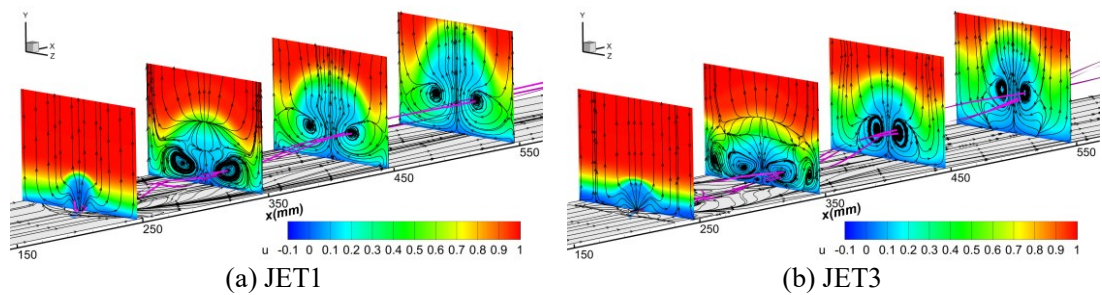

(a) JET1          (b) JET3

Fig. 39: Contours of $u$ and streamlines at the streamwise cross-sections (with $x = 250, 350, 450, 550mm$) under different AR with $\dot{m} = 1.5 g/s$ & $L_z = 45mm$.

Fig. 40 presents corresponding pressure contours and streamlines at the $z = 0$ cross-section and on the wall. As the AR increases, the reattachment line on the wall of the JET3 case initially approaches the spanwise boundaries and then shifts toward the center, indicating the influence of the air film's expansion. The flow structure on the $z = 0$ section further reveals that the low-pressure region becomes more confined following the JET3 injection; in addition to the existing separation and vortex, a large-scale source-type node emerges downstream.

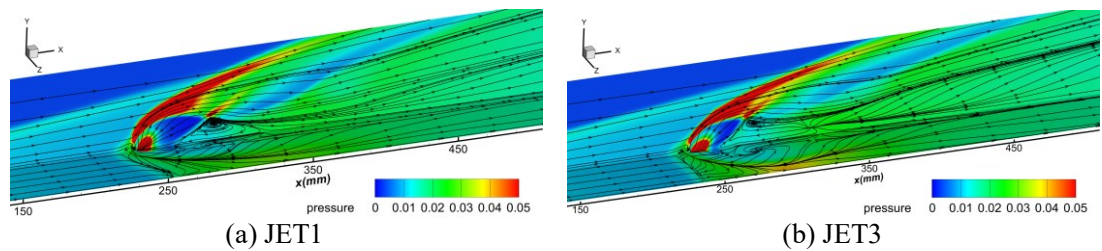

(a) JET1          (b) JET3

Fig. 40: Isobar contours and streamlines at the $z = 0$ cross-section and on the wall under different AR with $\dot{m} = 1.5 g/s$ and $L_z = 45 mm$.

(2) Effect on friction reduction performance

To quantitatively assess the influence of AR, Table 6 summarizes the $\eta_D$ of three holes, calculated using the same integration region as defined in Section 3. It is evident that $\eta_D$ markedly increases with increasing AR. Notably, JET3 achieves a value as high as 37.80%, nearly doubling that of JET2.

Table 6: Drag reduction efficiency under different aspect ratios of the holes.

| Hole | JET1 | JET2 | JET3 |
|---|---|---|---|
| $\eta_D$ (%) | 8.58 | 20.51 | 37.80 |

To analyze the underlying mechanism, Fig. 41 presents the contours of the $\Delta C_f$ of various AR cases with streamlines, as well as the isoline of $\Delta C_f = 0$. Upstream of the injections, the friction reduction in the JET1 case is found to be less pronounced compared to that of JET2. Downstream of the injection holes, the reattachment region in the JET1 case exhibits a substantially larger area of increased friction, suggesting that the streamwise vortices induce strong flow reattachment over an extended span. This results in flow downwash and increased friction. In contrast, the JET3 configuration exhibits a rather reduced extent and magnitude of friction increase, demonstrating superior friction reduction performance.

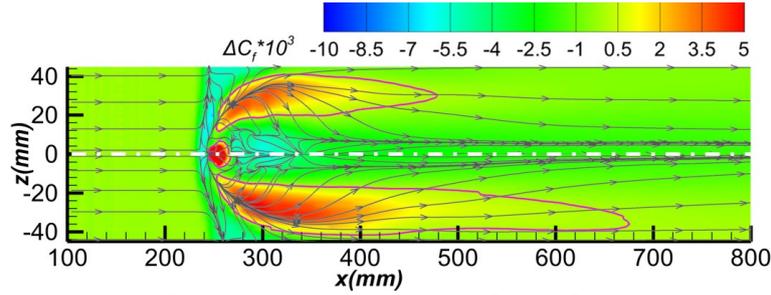

(a) $z \in [0,45]$ for the JET2 case, while $z \in [-45,0]$ for the JET1 case

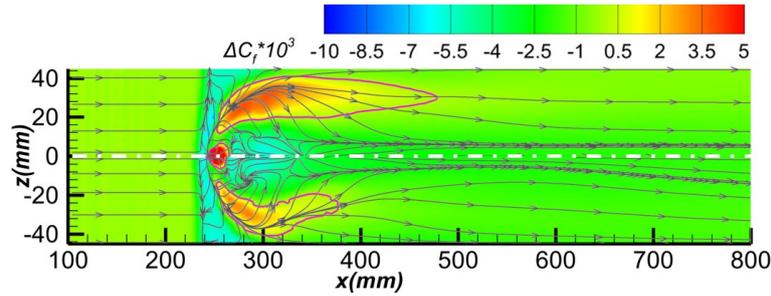

(b) $z \in [0,45]$ for the JET2 case, while $z \in [-45,0]$ for the JET3 case

Fig. 41: Contours of $\Delta C_f$ and streamlines under different ARs on the wall with $\dot{m} = 1.5 g/s$ & $L_z = 45 mm$ (dashed lines representing the boundary dividing different cases).

Fig. 42 depicts the $\bar{C}_f$ distributions (see Eq. (7)) with $x$ for holes with different ARs. As AR increases, the reduction in $\bar{C}_f$ upstream of the injection holes becomes more pronounced, the peak jump at the injection location decreases, and notably, the $\bar{C}_f$ downstream of the holes is significantly lowered.

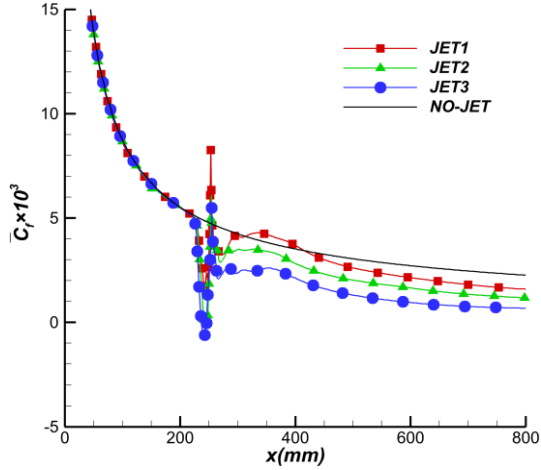 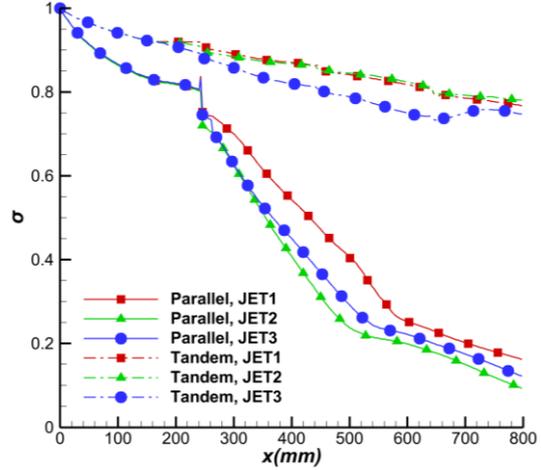

Fig. 42: Distributions of $\bar{C}_f$ with $x$ of injections with different AR with $\dot{m} = 1.5 g/s$ and $L_z = 45 mm$.

Fig. 43: Distributions of $\sigma$ with $x$ of injections with different AR with $\dot{m} = 1.5 g/s$ and $L_z = 45 mm$.

The contour distributions of $u$ and $\partial u/\partial y$ are checked at three spanwise sections at $z = 0, -L_z/2, -L_z$, respectively, similarly to those presented in Fig. 19 in Section 3.4. In general, similar characteristics are manifested in the results of the three JETs, which have been discussed in Section 3.3 in relation to the case of JET2. However, JET3 yields a relatively lower $u$ as well as $\partial u/\partial y$ near the wall, which favors friction reduction. The quantitative distributions of $u$ with $y$ have also been checked at $(x, z) = \{(325mm, 0), (625mm, 0)\}$ for the three JETs, and the velocity plateau is similar to those presented in Fig. 20 in Section 3.4. Scrutiny of the figure indicates that with increasing AR (e.g., JET3), the near wall low-speed region thickens and the value of plateau reduces somewhat, jointly yielding reduced shear on the wall. Furthermore, the PHs are also checked, finding similar distributions. Considering the discussions around JET2 presented in Section 3, the details are omitted here to save space.

As well as the benefit of friction reduction, the flow losses are also evaluated. Fig. 43 presents the corresponding distributions of $\sigma$ with $x$ for parallel injections of JETs1–3. The figure shows that the three JETs exhibit similar distributions with a relatively lower $\sigma$ for JET2 and JET3 following the injections. To save space, the results of JETs1–3 for the tandem injections with $\dot{m} = 0.3 g/s, L_h = 200 mm$ presented in Section 5.2 are also shown in Fig. 43.

## 5.2 Effect of aspect ratio in tandem injection

(1) Effect on flow structures

Fig. 44 also presents the contours of $u$ and streamlines at six streamwise sections for the JET1 and JET3 cases, which are similar to those of JET2 presented in Section 4.2. Consistent with the observations from Fig. 39, Fig. 44 shows that as AR increases, (1) the heights of the primary vortices downstream of each injection hole decrease along with increasing spanwise scale as indicated by the sections between the holes, implying decreased flow transportation; (2) the fluid in the transition with moderate $u$ (depicted in green) near spanwise boundaries is less entrained toward the wall by the vortices; (3) the bulge formed by the lifted flow with low $u$ from the two vortices decreases in height. The spanwise extent of the near wall region with low-speed $u$ expands, and lateral coverage of the air film broadens, which is conducive to enhancing friction reduction.

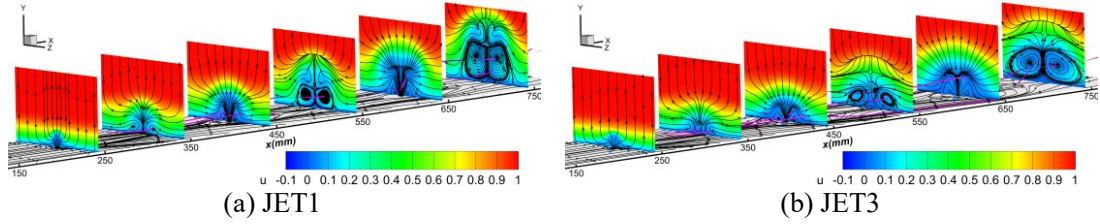

(a) JET1  (b) JET3

Fig. 44: Contours of $u$ and streamlines at the streamwise cross-sections (with $x = 250, 350, 450, 550, 650, 750mm$) under different AR and with $\dot{m} = 0.3g/s$ and $L_h = 200mm$.

Fig. 45 presents corresponding pressure contours and streamlines at the $z = 0$ section and on the wall for the JET1 and JET3 cases. With increasing AR, intact separation bubbles are seen to form at the $z = 0$ section downstream of Holes-1 and -2, while the recirculation zone downstream of Hole-3 expands; the spanwise extent of the wall reattachment line broadens, indicating an enlarged lateral influence of the air film.

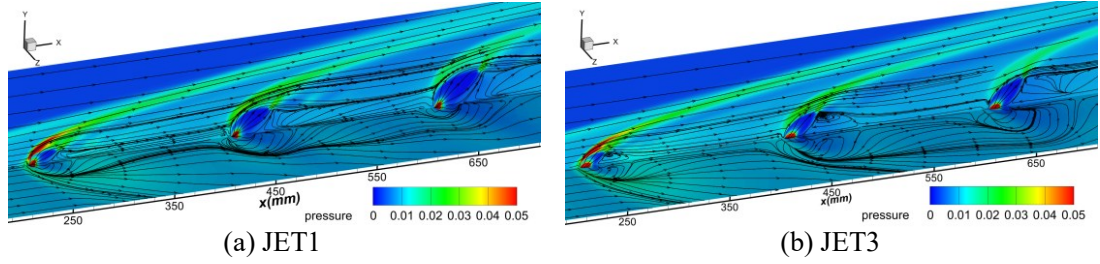

(a) JET1  (b) JET3

Fig. 45: Isobar contours and streamlines at the $z = 0$ cross-section and on the wall under different AR and with $\dot{m} = 0.3g/s$ and $L_h = 200mm$.

(2) Effect on friction reduction performance

To quantitatively compare the influence of AR, Table 7 presents $\eta_D$ for three holes, with an integration region identical to that in Section 4. The results reveal that $\eta_D$ increases significantly with increasing AR. Notably, the JET3 configuration achieves the highest $\eta_D$ of 45.95%, nearly doubling that of the JET1 case.

Table 7: Drag reduction efficiency under different aspect ratios of the holes.

| Hole | JET1 | JET2 | JET3 |
| --- | --- | --- | --- |
| $\eta_D(\%)$ | 24.91 | 36.94 | 45.95 |

To explore the underlying mechanisms, Fig. 46 depicts the contours of the $\Delta C_f$ of three AR cases as well as the isoline of $\Delta C_f = 0$, along with corresponding streamlines. It can be observed that with increasing AR, the spanwise extent of the friction reduction zones downstream of each hole expands; the regions with $\Delta C_f \geq 0$ adjacent to Hole-1 significantly decrease, while those adjacent to Holes-2 and -3 gradually disappear, demonstrating enhanced friction reduction performance.

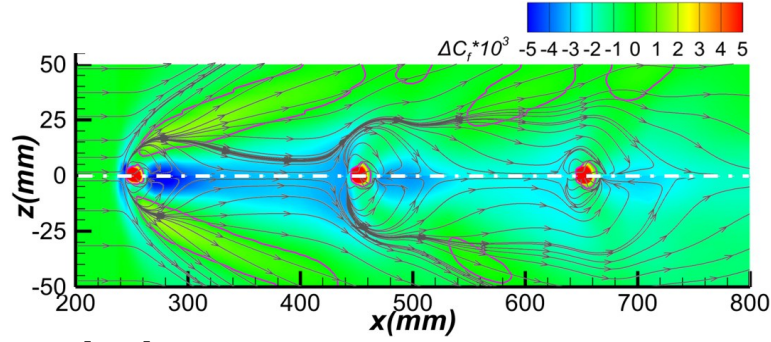

(a) $z \in [0,50]$ for the JET1 case, while $z \in [-50,0]$ for the JET2 case

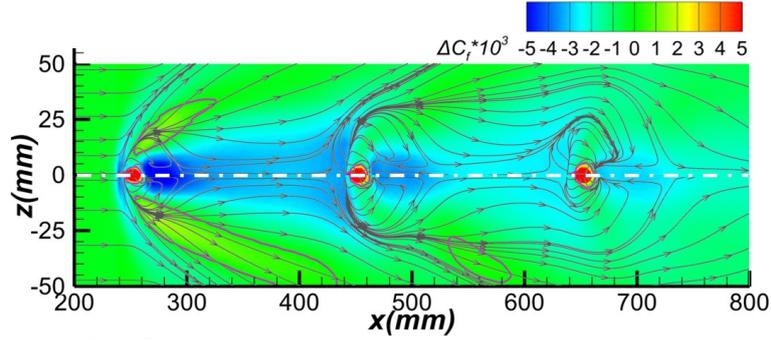

(b) $z \in [0,50]$ for the JET3 case, while $z \in [-50,0]$ for the JET2 case

Fig. 46: Contours of $\Delta C_f$ and streamlines under different AR on the wall and with $\dot{m} = 0.3 g/s$ and $L_h = 200mm$ (dashed lines representing the boundary dividing different cases).

Fig. 47 illustrates the streamwise distributions of $\bar{C}_f$ for three AR cases. The results indicate that as AR increases, the $\bar{C}_f$ downstream of each injection decreases, and the peak of its abrupt change at the injection is reduced.

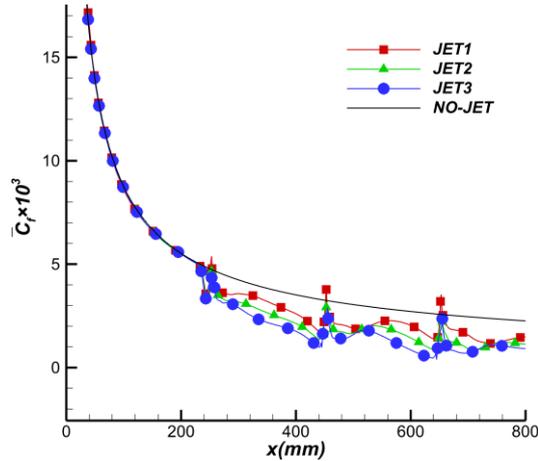

Fig. 47: Distributions of $\bar{C}_f$ with $x$ under different AR and with $\dot{m} = 0.3 g/s$ and $L_h = 200mm$.

The contours of $u$ and $\partial u/\partial y$ are examined at three spanwise sections with $z = 0, -L_z/2, -L_z$ respectively, as were those in Fig. 36 presented in Section 4.4. The distributions are substantially similar among the three JETs, and have been discussed in the case of JET2 in Section 4.3. Differences included JET3 engendering a lower $u$ and decreasing $\partial u/\partial y$ near the wall, implying an improved spanwise coverage of the air film. In addition, quantitative checks are made of the normal distributions of the $u$ of the three JETs at two positions, namely the middle position between Holes-1 and -2 and the one where $(x, z) = (750mm, 0)$. The results show that the at the first position, the three JETs exhibit similar distributions, while at the second position, increasing AR results in a reduction in the quasi-plateau velocity, which is favorable for decreasing the near wall velocity gradient. Considering the discussions on JET2 have been presented in Section 4, the details are omitted here to save space. As noted in Section 5.1, the flow losses are evaluated as well as the friction reduction, with results of the three JETs presented in Fig. 43. It can be observed that in the tandem injections, JET3, having the largest AR, manifests relatively larger losses.

In summary, this section investigated the effect of the AR of a given hole on friction reduction. The results demonstrate that increasing AR leads to enhanced friction reduction, which has potential significance for practical engineering. The underlying mechanisms are similar for both parallel and tandem injections. When the AR is small, the downstream primary vortices exhibit enhanced transport capability, entraining higher velocity fluid closer to the wall, which weakens the friction reduction compared to larger AR configurations. Given that variations in AR have relatively less influence on flow losses, configurations with larger ARs are recommended to maximize friction reduction. It is worth noting that Zhang et al. [34] compared the cooling performance of injection holes with ARs of 0.5, 1, and 2 under low-speed inflow conditions (with inlet velocity of 20 m/s). Their findings suggested that superior cooling effectiveness would be achieved with smaller ARs, contrasting with the present results under hypersonic conditions, where larger ARs yield better friction reduction.

## 6 Conclusions

This study conducted numerical simulations based on the 3-D N–S equations using the WENO3-$Z_{ES3}$ scheme and an EGM. The friction reduction caused by air films generated by parallel and tandem injections on a flat plate was investigated under $Ma_\infty = 15.0$ at an altitude of $60 km$. The influence of $\dot{m}$, $L_z$, $L_h$, and the AR of the holes were discussed in detail. The conclusions are drawn as follows:

(1) Due to the exertion of the air film under hypersonic and high-altitude conditions, the freestream is lifted, leading to the formation of a separation zone upstream and a recirculation zone downstream of the injection. A distinct low-speed region appears near the wall downstream of the injection, accompanied by a pair of dominant counter-rotating vortices. In the wall-normal direction, a plateau forms in the $u$ profile, with a value significantly lower than $U_\infty$. This results in a reduced velocity gradient near the wall, which contributes to friction reduction. While the air film reduces friction within its coverage area, conversely, the flow reattachment downstream of the injection can increase friction locally.

(2) For both parallel and tandem injections, increasing the $\dot{m}$ leads to a higher PH of the air film, an expanded wall-normal extent of the low speed flows, reduced velocity gradients near the wall, and increased spanwise coverage of the film. However, the enlarged local regions of increased friction reported in (1) may limit the overall improvement in friction reduction. Meanwhile, the higher $\dot{m}$ is generally accompanied by increased total pressure loss.

(3) With increasing hole spacing, in parallel injections, the PH of the air film is observed to decrease, and the spanwise coverage relative to the plate width reduces. The uniformity of the spanwise distribution of $u$ near the wall and friction reduction decrease, although total pressure loss also reduces. Meanwhile, tandem injections exhibit smaller variations in flow structure, friction reduction, and flow loss with changing $L_h$. The optimal selection of $\dot{m}$ and hole spacing requires a balance between friction reduction and flow loss.

(4) The investigations of the effect of the AR of the holes show that increasing the AR reduces

the scale of primary vortices and weakens their ability to entrain high-speed fluid toward the wall. This enhances the spanwise uniformity of the $u$ distribution, reduces near wall velocity gradients, and enlarges the spanwise extent of the low-speed region. As a result, the area of local friction increase is effectively diminished, and overall friction reduction is markedly improved. Notably, under hypersonic flow conditions, the observed trends in the friction reduction associated with AR are the reverse of those reported by low-speed studies, where holes with smaller ARs typically yield better performance. Hopefully, this understanding will be helpful for optimizing drag control in engineering.

**Acknowledgements**

It is grateful to Guozhuo Tan for his preliminary work on air film which motivates this study.

**References**

[1] Y. Zhang, Y. Zhang, J. Xie, et al. Study of viscous interaction effect model for typical hypersonic wing-body figuration. Acta Aerodynamica Sinica, 2017, 35(02): 186-191.

[2] L. Pan, H. Hao, Z. Yao, et al. Current status of research on reducing drag and cooling of high-speed aircraft. Advances in Mechanics, 2023, 53(04): 793-818.

[3] X. Sun, W. Huang, M. Ou, et al. A survey on numerical simulations of drag and heat reduction mechanism in supersonic/hypersonic flows. Chinese Journal of Aeronautics, 2019, 32(04): 771-784.

[4] G. Zhu, S. Yao, Y. Duan. Research progress and some issues in engineering application of flow control technology for drag reduction and heat reduction of high-speed flight vehicles. Acta Aeronautica et Astronautica Sinica, 2023, 44(15):1-17.

[5] H. Zhang, J. Huang, S. Gao. Numerical simulation of hypersonic flow over axisymmetric spiked body. Acta Aeronautica et Astronautica Sinica, 1994, 15(05):519-525.

[6] S. Saravanan, G. Jagadeesh, K. P. J. Reddy. Investigation of missile-shaped body with forward-facing cavity at Mach 8. Journal of Spacecraft and Rockets, 2009, 46(03): 577-591.

[7] H. Lu, W. Liu. Investigation on thermal protection efficiency of hypersonic vehicle nose with forward-facing cavity. Journal of Astronautics, 2012,33(08):1013-1018.

[8] H. Lu, W. Liu. Cooling efficiency investigation of forward-facing cavity and opposing jet combinatorial thermal protection system. Acta Physica Sinica, 2012, 61(06): 372-377.

[9] Y. Zhou, Z. Luo, L. Wang, et al. Plasma synthetic jet actuator for flow control: Review. Acta Aeronautica et Astronautica Sinica, 2022, 43(03): 98-140.

[10] Z. Ma, Z. Luo, A. Zhao, et al. Reverse jet characteristics of plasma synthetic jet in hypersonic flow field. Acta Aeronautica et Astronautica Sinica, 2022, 43(S2): 195-206.

[11] N. Sahoo, V. Kulkarni, S. Saravanan, et al. Film cooling effectiveness on a large angle blunt cone flying at hypersonic speed. Physics of Fluids, 2005, 17(03): 036102.

[12] L. Wang, Y. Wang, Z. Luo, et al. Study on the drag reduction characteristics of opposing jet on hypersonic lifting body configuration. Journal of Astronautics, 2024, 45(06): 881-892.

[13] J. Chang, W. Bao, D. Yu, et al. Effects of wall cooling on performance parameters of hypersonic inlets. Acta Astronautica, 2009, 65(3-4): 467-476.

[14] L. Tang, P. Li, L. Zhou. Review on liquid film cooling of liquid rocket engine. Journal of Rocket Propulsion, 2020, 46(01): 1-12.

[15] J. Zuo, S. Zhang, J. Wei, et al. Review on fuel supersonic film thermal protection/drag reduction cooperative technology for internal flow of hypersonic vehicles. Journal of Propulsion Technology, 2025, 46(01):6-24.

[16] B. Chen, S. Yi. Cooling characteristics for supersonic film on flat/curved surfaces in hypersonic

mainstream. Journal of National University of Defense Technology, 2025, 47(01): 126-135.

[17] Z. Hu, Y. Xing, F. Zhong. Flow and heat transfer characteristics of film cooling on hypersonic nozzles. Journal of Propulsion Technology, 2023, 44(12): 115-123.

[18] F. Mo, W. Su, Z. Gao, et al. Numerical investigations of the slot blowing technique on the hypersonic vehicle for drag reduction. Aerospace Science and Technology, 2022, 121: 107372.

[19] J. Lin, Q. Wang, Y. Zhao, et al. An experimental investigation of supersonic conical cooling films with angles of attack. Physics of Fluids, 2024, 36(10): 106101.

[20] J. Lin, Q. Wang, Y. Zhao, et al. An experimental study of supersonic conical cooling films subjected to different ratios of static pressure. Physics of Fluids, 2024, 36(08): 086122.

[21] X. Zhao, S. Yi, Q. Mi, et al. Drag reduction of a hypersonic cone with supersonic cooling film. Journal of Thermophysics and Heat Transfer, 2022, 36(01): 221-225.

[22] X. Sun, H. Ding, M. Liu, et al. Experimental investigation on supersonic film cooling of hypersonic optical dome under different nozzle pressure ratios. Aerospace Science and Technology, 2023, 140: 108488.

[23] X. Sun, H. Ding, S. Yi, et al. Research on supersonic film cooling of hypersonic optical window under different nozzle pressure ratios. Physics of Fluids, 2024, 36(10): 106131.

[24] M. Hombsch, H. Olivier. Film cooling in laminar and turbulent supersonic flows. Journal of Spacecraft and Rockets, 2013, 50(04): 742-753.

[25] A. S. Pudsey, R. R. Boyce, V. Wheatley. Hypersonic viscous drag reduction via multiport hole injector arrays. Journal of Propulsion and Power, 2013, 29(05): 1087-1096.

[26] M. A. Keller, M. J. Kloker, H. Olivier. Influence of cooling-gas properties on film-cooling effectiveness in supersonic flow. Journal of Spacecraft and Rockets, 2015, 52(05): 1443-1455.

[27] J. Zhou. Numerical investigation on supersonic film cooling performance with discrete film holes. AIAA Journal, 2023, 61(01): 48-62.

[28] P. Kerth, L. M. Le Page, S. Wylie et al. Displacement of hypersonic boundary layer instability and turbulence through transpiration cooling. Physics of Fluids, 2024, 36(03): 034102.

[29] S. Dai, H. Ma, Z. Xu, et al. Numerical study of discrete film cooling near the rudder shaft of a hypersonic air fin model. Numerical Heat Transfer; Part A: Applications, 2024, 85(05): 679-701.

[30] S. Xiang, S. Shang, Z. Shen, et al. Research progress and development direction of hypersonic film cooling technology. Aerospace Materials & Technology, 2020, 50(03): 1-10.

[31] J. Mao, G. Qu, Z. Gao. Numerical investigation of heat and drag reduction by discrete microholes film in hypersonic flow. Journal of Beijing University of Aeronautics and Astronautics.2024. doi: 10.13700/j.bh.1001-5965.2024.0443.

[32] Q. Li, P. Yan, X. Huang, et al. Improvements to enhance robustness of third-order scale-independent WENO-Z schemes. Advances in Applied Mathematics and Mechanics, 2025 17(02): 373-406.

[33] J. C. Tannehill, R. A. Mohling. Development of equilibrium air computer programs suitable for numerical computation using time-dependent or shock-capturing methods. NASA CR-2470, 1974.

[34] G. Zhang, J. Liu, B. Sundén, et al. Comparative study on the adiabatic film cooling performances with elliptical or super-elliptical holes of various length-to-width ratios. International Journal of Thermal Sciences, 2020, 153: 106360.